\documentclass[a4paper,12pt]{article}
\usepackage{graphicx}
\usepackage{amssymb, amsmath}
\begin{document}
\def\lsim{\:\raisebox{-0.5ex}{$\stackrel{\textstyle<}{\sim}$}\:}
\def\gsim{\:\raisebox{-0.5ex}{$\stackrel{\textstyle>}{\sim}$}\:}
\def\mET{E_T \hspace{-1.1em}/\;\:}
\def\mpT{p_T \hspace{-1em}/\;\:}
\def\sh{\mbox{\texttt h}}
\def\et{\mbox{${E\!\!\!/}_T$}} 
\def\rpv{\mbox{${R\!\!\!/}_p$}} 
\def\eplem{\mbox{$e^+e^-$}}
\def\epem{e^+e^-}
\def\emem{e^-e^-}
\newcommand{\cd}{{{\cal D}}}
\newcommand{\cv}{{{\cal V}}}
\newcommand{\cl}{{{\cal L}}}
\def\gamgam{\mbox{$\gamma \gamma$}}     
\def\rts{\mbox{$\sqrt{s}$}}   
\def\ma{\mbox{$m_A$}}  
\def\B{{\mbox{\boldmath $B$}}} 
\def\W{{\mbox{\boldmath $W$}}} 
\def\Journal#1#2#3#4{{#1} {\bf #2}, #3 (#4)}
\def\NCA{\em Nuovo Cimento}
\def\NIM{\em Nucl. Instrum. Methods}
\def\NIMA{{\em Nucl. Instrum. Methods} A}
\def\NPB{{\em Nucl. Phys.} B}
\def\PLB{{\em Phys. Lett.}  B}
\def\PR{\em Phys. Reports}
\def\EUJ{{\em Eur. Phys. J.} C}
\def\PRL{\em Phys. Rev. Lett.}
\def\PRD{{\em Phys. Rev.} D}
\def\ZPC{{\em Z. Phys.} C}
\def\PR{{\em Phys. Reports}}
\def\RMP{\em Rev. of Modern Physics}
\def\JHEP{\em Journal of High Energy Physics}
\def\pramana{\em Pramana}
\def\bedes{\begin{description}} 
\def\edes{\end{description}}
\def\eg{ {\em e.g.}~}
\def\N0{\widetilde \chi^0}
\newcommand{\beqn}{\begin{eqnarray}}
\newcommand{\eeqn}{\end{eqnarray}}
\newcommand{\tr}{{\rm Tr}} 
\newcommand{\be}{\begin{equation}}
\newcommand{\beq}{\begin{equation}}
\newcommand{\eeq}{\end{equation}}
\newcommand{\ee}{\end{equation}}                   
\newcommand{\bea}{\begin{eqnarray}}
\newcommand{\eea}{\end{eqnarray}}             
\def\ie{ {\em i.e.}} 

                         \def\Chip{\widetilde \chi^+}
                         \def\Chim{\widetilde \chi^-}
                         \def\Chipm{\widetilde \chi^\pm}
                         \def\Chimp{\widetilde \chi^\mp}
                         \def\sq{\widetilde q}
                         \def\su{\widetilde u}
                         \def\sd{\widetilde d}
                         \def\sc{\widetilde c}
                         \def\ss{\widetilde s}
                         \def\st{\widetilde t}
                         \def\sb{\widetilde b}
\def\sl{\widetilde \ell}
\def\sel{\widetilde e}
\def\snu{\widetilde \nu}
\def\smu{\widetilde \mu}
\def\stau{\widetilde \tau}
\def\glu{\widetilde g}
\def\tm{\widetilde m}

\thispagestyle{empty}
\begin{flushright}                               
                                                    hep-ph/0205114  \\
                                                    IISc/CTS/17-01   \\ 
\end{flushright}
\begin{center}

{\LARGE\bf
Physics Potential of the Next Generation Colliders.}\footnote {Invited 
article for the special issue  on  High Energy Physics  of the Indian Journal 
of Physics on the occasion of its Platinum Jubilee.}
\vskip 25pt
         R.M. Godbole

\bigskip
 Centre for Theoretical Studies,\\
Indian Institute of Science, Bangalore,
560012, India.\\
 E-mail: rohini@cts.iisc.ernet.in\\

\bigskip
           Abstract
\end{center}

\begin{quotation}
\noindent
In this article I summarize some aspects of the current status of the field
of high energy physics and discuss how the next generation of high energy
colliders will aid in furthering our basic understanding of elementary 
particles and interactions among them, by shedding light on the mechanism 
for the spontaneous breakdown of the Electroweak Symmetry.
\end{quotation}
\newpage

\section{Introduction:}

 Particle physics is at an extremely interesting juncture at present. The
theoretical developments of the last 50 years have now seen establishment
of quantum gauge field theories as the paradigm for the description of
fundamental particles and interactions among them. The Standard Model of
particle physics (SM) which provides a description of particle interactions
in terms of a Quantum Gauge Field Theory with $SU(3)_{C}$ X $SU(2)_{L}$ X 
$U(1)$  gauge
invariance, has been shown to describe all the experimental observations in
the area of electromagnetic, weak and strong interactions of quarks and
leptons. The predictions of the Electroweak (EW) theory have been tested to an
unprecedented accuracy. These predictions involve effects of loop
corrections, which can be calculated in a consistent way only for a
renormalizable theory and the theories are guaranteed to be renormalizable,
if they are gauge invariant. Naively, the gauge invariance is
guarunteed only if the corresponding gauge boson is massless. The massless
$\gamma$ is an example.
 Developement of a unified theory of electromagnetic and weak interactions
in terms of a gauge invariant quantum field theory (QFT) took place in the 70s and
80s. The theoretical cornerstone of these developments was the proof that
these theories are renormalizable even in the presence of nonzero masses of the
corresponding gauge bosons W and Z; the so called spontaneous breakdown of the
gauge symmetry. The first experimental proof in
favour of the EW theory came in the form of the discovery of neutral
current interactions in 1973, whereas the first direct experimental
observationxs of the massive gauge bosons came in 1983 at the 
S$\bar{p}$pS collider.
The measurement of the masses of the W and Z bosons in these experiments,
their agreement with the predictions of the Glashow, Salam and 
Weinberg (GSW) model in terms of a single parameter 
$sin^{2}\theta_{W}$ determinded experimentally from a variety of data and 
the verification of the relation $\rho$ = $\frac{m^{2}_{W}}{m^{2}_{Z}
cos^{2} \theta_{W}} = 1$  were the various milestones in the establishment 
of the GSW model as {\it the} correct theory of EW interactions as 
an SU(2) X U(1) gauge theory at the tree level.
However, the correctness of this theory at the loop level was proved
conclusively only by the spectacular agreement of the value of 
$m_{t}$ obtained from {\bf direct} observations of the top quark at 
the $\bar{p}$p collider Tevatron ($m_{t} =174.3 \pm 5.1 $ GeV) with the one 
obtained {\bf indirectly} from the precision measurements of the properties 
of the Z boson at the $e^{+}e^{-}$ LEP and SLC colliders 
($m_{t}= 180.5 \pm 10.0 $ GeV).  Direct observations of the effects of 
the trilinear WWZ coupling, reflecting the nonabelian nature of the 
$SU(2)_{L}$ X U(1) gauge theory through the direct measurement of the 
energy dependence of $\sigma(e^{+}e^{-} \to W^{+}W^{-})$ at the second stage 
of $e^+e^-$ collider LEP (LEP2)  also
was an important milestone. As a result of a variety of high precision
measurements in the EW processes, the predictions of the
SM as a QFT with $SU(3)_{C}$ X $SU(2)_{L}$  X $ U(1)$ gauge invariance, 
have now been tested to an accuracy of 1 part in $10^{6}$. Fig. 1 
from~\cite{LEPC} shows the
\begin{figure}[htb]                
       \centerline{
      \includegraphics*[scale=0.40]{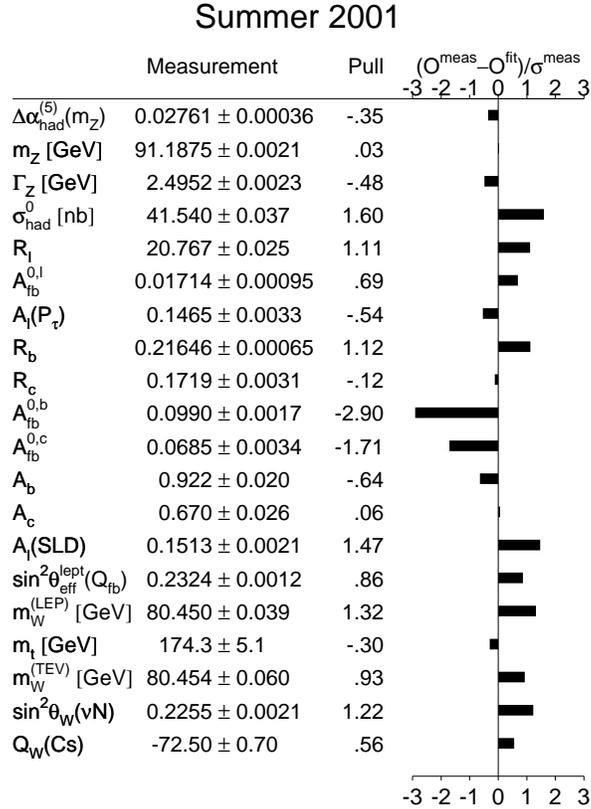}
}
\caption{\em  Precision measurements at the \eplem\ colliders LEP/SLC and `pull'
\label{fig1}}           
\end{figure}
precision measurements of a large number of observables at LEP/SLC alongwith
their best fit values to the SM in terms of $m_{Z}$, $\alpha_{em}$ and the 
nuclear $\beta$ decay,  Fermi coupling constant 
$G_{F}$. The third column shows the `pull', i.e., the difference between 
the SM fit value and the measurements in terms of standard deviation 
error of the measurement.  It is interesting that due to the very
high precision of these measurements the $\chi^2 /dof$ is $22.9/15$ in
spite of the very good quality of fits provided by the SM. The possible 
precision now even allows extracting indirect
information on the Higgs mass, just the same way the earlier LEP
measurements at the Z-pole, gave the top mass indirectly. In spite of this
impressive success of the SM, in describing every piece of measurement in
high energy experiments so far (with the possible exception of the
$(g-2)_{\mu}$ measurement~\cite{gm2}), the complete truth of the SM 
will not be established {\it satisfactorily} until one finds `direct' 
evidence for the Higgs scalar. In
addition to this we should also remember that although the beautiful
experiments at HERA and Tevatron, have confirmed the predictions of
perturbative QCD in the domain of large $Q^{2}$, there has been no `direct'
observation of the `unconfined' quarks and gluons.
 In the next section I first discuss why the idea of the Higgs boson forms
the cornerstone of the SM model. Further I point out the arguments which
show that either a Higgs scalar should exist in the mass range indicated by
the current precision measurements~\cite{LEPC}, or there should exist (with a
very high probability) some alternative physics at $\sim$ TeV scale which would
provide us with a clue to an understanding of the phenomenon of spontaneous
breakdown of the EW symmetry. In the next section I will discuss what light 
will the
experiments at the Tevatron in near future be able to shed on the problem. 
I will not, however, be able to survey what information the currently 
running experiment at Relativistic Heavy Ion Collider (RHIC) will provide 
about the theory of strong interactions, QCD,  in a domain not accessible 
to perturbative quantum computation. I will end by  discussing the role 
that the future colliders will play in unravelling this last knot in our 
understanding of fundamental particles and interactions among them. These 
future colliders consist of the pp collider, Large Hadron Collider (LHC) 
which is supposed to go in action in 2007 and will have a total c.m. 
energy of 18 TeV and the higher energy $e^{+}e^{-}$ linear colliders
with a c.m. energy between $350$ GeV and $1000$ GeV,  which 
are now in the planning stages.

\section{The spontaneous breakdown of EW symmetry and new physics at the TeV
scale}

Requiring that cross-sections involving weakly interacting particles 
satisfy unitarity, {\it viz.}, they rise slower than $\log^2 (s)$  at
high energies, played a very important role in the theoretical development 
of the EW theories. Initially the existence of intermediate vector bosons 
itself was postulated to avoid `bad' high energy rise of the neutrino 
cross-section. Note that possible
violations of unitarity for the cross-section $\bar{\nu_{\mu}}$ $\mu \to
\bar{\nu_{e}}$ e for E $<$ $G_{F}^{-1/2}$ $\sim$ 300 GeV is cured in 
reality by a $W^{-}$ boson of a much lower mass $\sim$ 80 GeV.
Further, the need for a nonabelian coupling as well as the existence of a scalar
with couplings to fermions/gauge bosons proportional to their masses can
be inferred ~\cite{Cornwall} from simply demanding good high energy behaviour 
of the cross-section $e^{+}e^{-} \rightarrow W^{+}W^{-}$. Left panel in Fig. 2 
shows the data on $W^{+}W^{-}$ production cross-section from LEP2. The 
data show the flattening of $\sigma$ ($e^{+}e^{-} \to W^{+}W^{-}$) at 
high energies clearly demonstrating the existence of the
ZWW vertex as well as that of the interference between the t-channel $\nu_{e}$
exchange and s-channel $\gamma^{*}/Z^{*}$ exchange diagrams. These diagrams are
indicated in the figure in the panel on the r.h.s. which shows the behaviour
of the same cross-section at much higher energies along with the
contributions of the individual diagrams~\cite{Fawzi}. This figure essentially
shows how the measurements of this cross-section at higher energies will test 
this feature of the SM even more accurately.
\begin{figure*}[htbp]
\begin{center}
\includegraphics{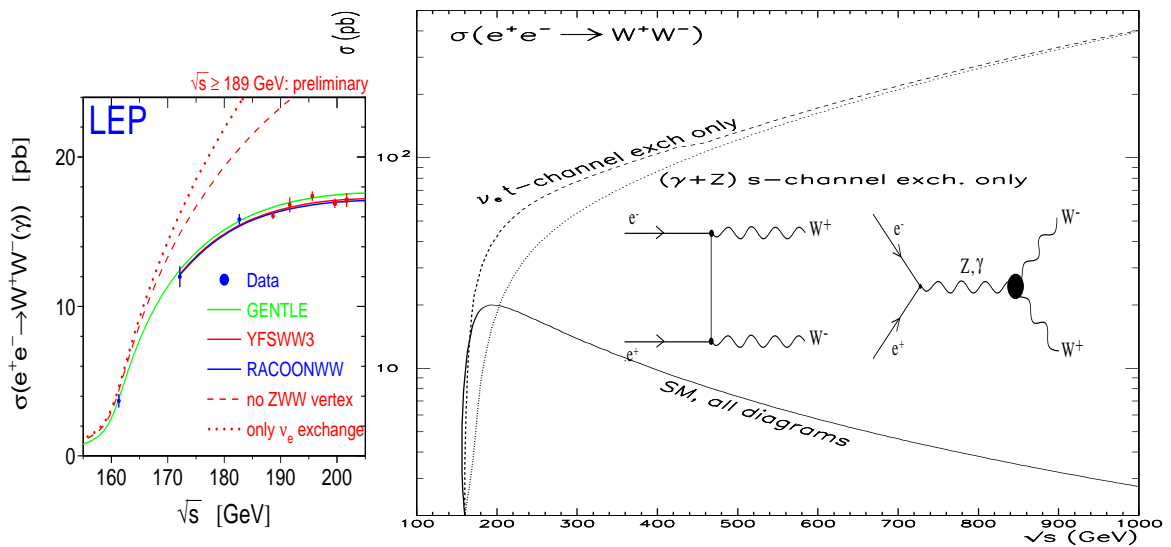}
\caption{\label{eetoww}{\em The small insert shows the latest data
of the $W^+W^-$ cross section at LEP2\cite{LEPC}. The main
figure shows the behaviour of the same cross section at much
higher energies and the contribution of each channel\/.}}
\end{center}
\end{figure*}
As mentioned in the introduction, one way to formulate a consistent,
renormalizable gauge field theory, is via the mechanism of spontaneous
symmetry breakdown of the $SU(2)_{L} X U(1)$ gauge symmetry, whose 
existence has been proved
incontrovertially by the wealth of precision measurements. Though the
mechanism predicts the  Higgs scalar and gives the couplings of
this scalar in terms of gauge couplings, its mass is not predicted.

The above discussion mentions the connection between unitarity
and existence of a Higgs scalar. Demanding $s$ wave unitarity of the process
$W^{+}W^{-} \to W^{+}W^{-}$ can actually give an upper limit on 
the mass of the Higgs.
Without the Higgs exchange diagram one can show that the amplitude for the
process $W^{+}_{L}W^{-}_{L} \to W^{+}_{L}W^{-}_{L}$ will grow with 
energy and violate unitarity for $\sqrt{s_{WW}} \geq$ 1.2 TeV, implying 
thereby that some physics beyond the gauge
bosons alone, is required somewhere below that scale to tame this bad high
energy behaviour. The addition of Higgs exchange diagrams tames the high
energy behaviour somewhat. Demanding perturbative unitarity for the J=0
partial wave amplitudes for  $W^{+}_{L}W^{-}_{L} \to W^{+}_{L}W^{-}_{L}$ 
gives a limit on the Higgs mass $m_{h} \leq \sim$ 700 GeV ~\cite{Lee-Thacker}. 
These arguments tell us that 
just a demand of unitarity implies that some new physics other than the gauge
bosons must exist at a TeV scale. Specializing to the case of the SM where
spontaneous symmetry breakdown happens via Higgs mechanism, the scale of the
new physics beyond just gauge bosons and fermions is lowered to 700 GeV.

\begin{figure}[htb]                
       \centerline{
      \includegraphics*[scale=0.53]{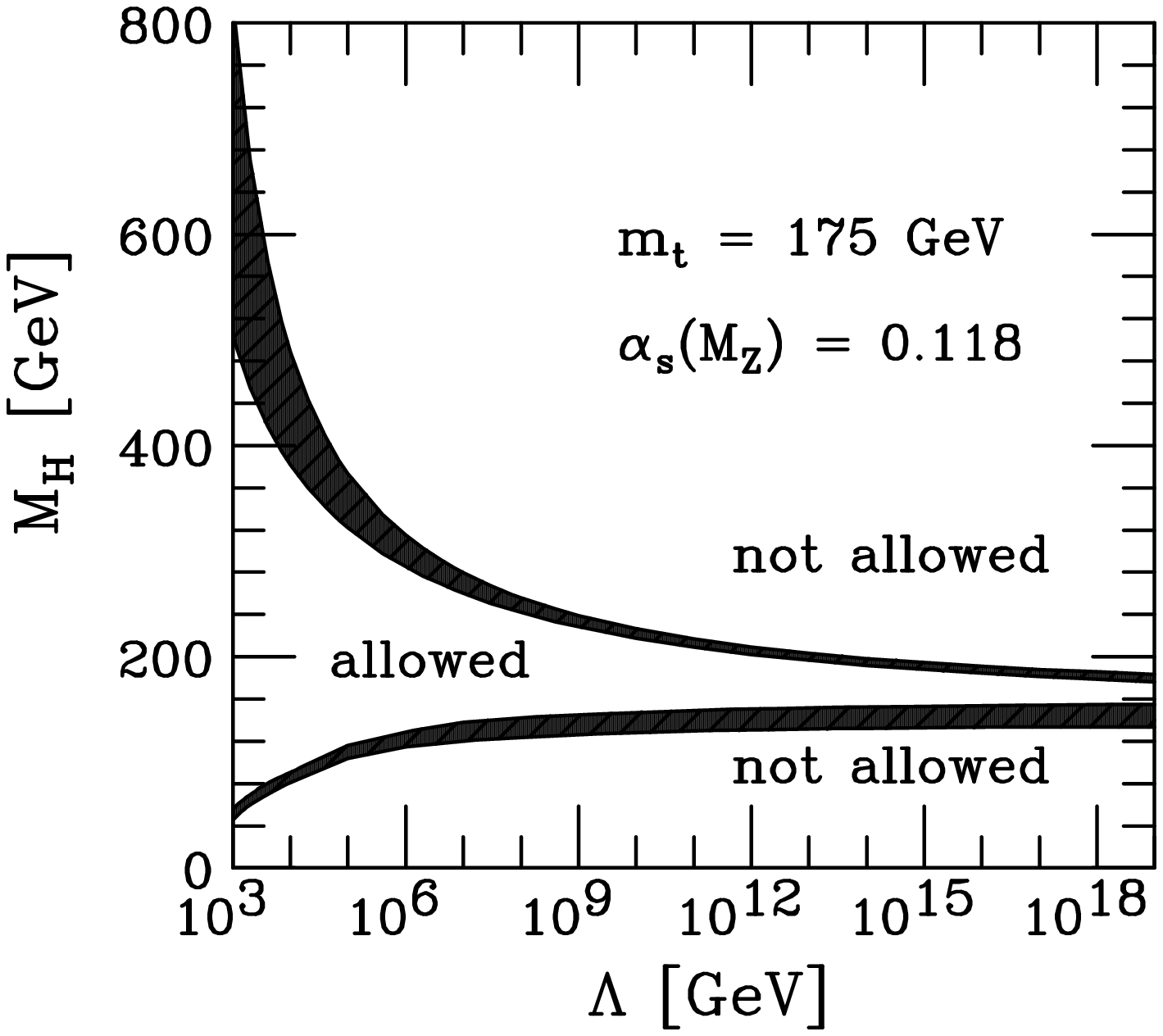}
      \includegraphics*[scale=0.35]{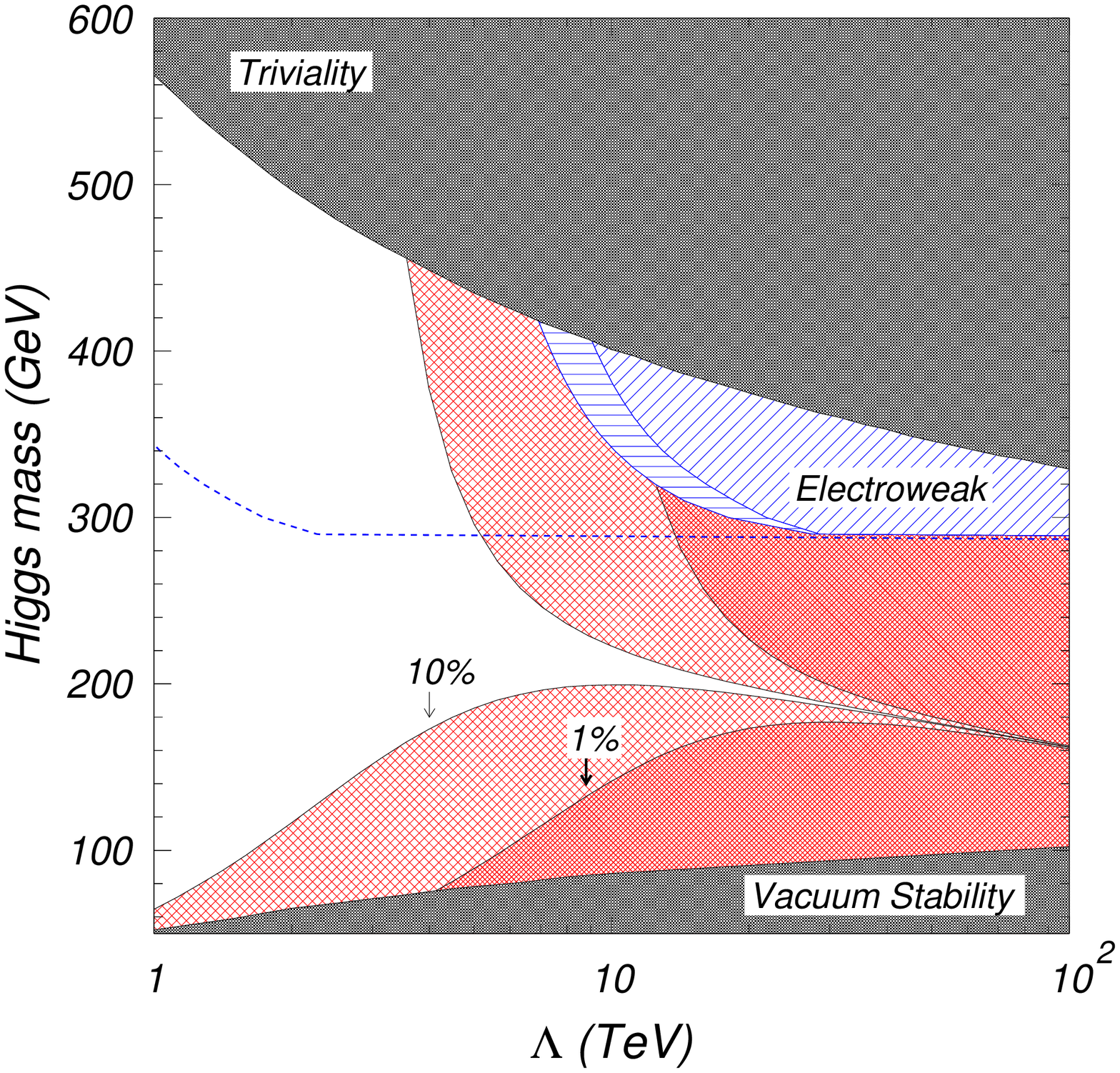}
  }
\caption{\em  Limits on the Higgs mass in the SM and 
beyond\protect\cite{hambye,murayama}.
\label{fig3}}           
\end{figure}

Within the framework of the SM, Higgs mass $m_{h}$ is also restricted by
considerations of vacuum stability as well as that of triviality of a pure
$\Phi^{4}$ field theory. The demand that the Landau pole in the self coupling
$\lambda$, lies above a scale $\Lambda$ puts a limit on the value of 
$\lambda$ at the EW scale which in turn limits the Higgs mass $m_{h}$.
This requirement essentially means that the Standard Model is a consistent 
theory upto a scale $\Lambda$  and no other physics need exist upto that scale.
The left panel in Fig.\ref{fig3} taken from~\cite{hambye}  shows the region 
in $m_{h}$ - $\Lambda$ plane that is allowed by these considerations, 
the lower limit coming from demand of vacuum stability. Note that 
for $\Lambda$ $\sim$ 1 TeV, the limit on $m_{H}$ is $\sim$ 800 GeV,
completely consistent with the unitarity argument presented above.

As mentioned in the introduction, the knowledge on $m_{t}$ from direct
observation and measurements at the Tevatron, now allows one to restrict
$m_{h}$ by considering the dependence of the EW loop corrections to 
the various EW observables
\begin{figure}[htb]                
       \centerline{
      \includegraphics*[scale=0.35]{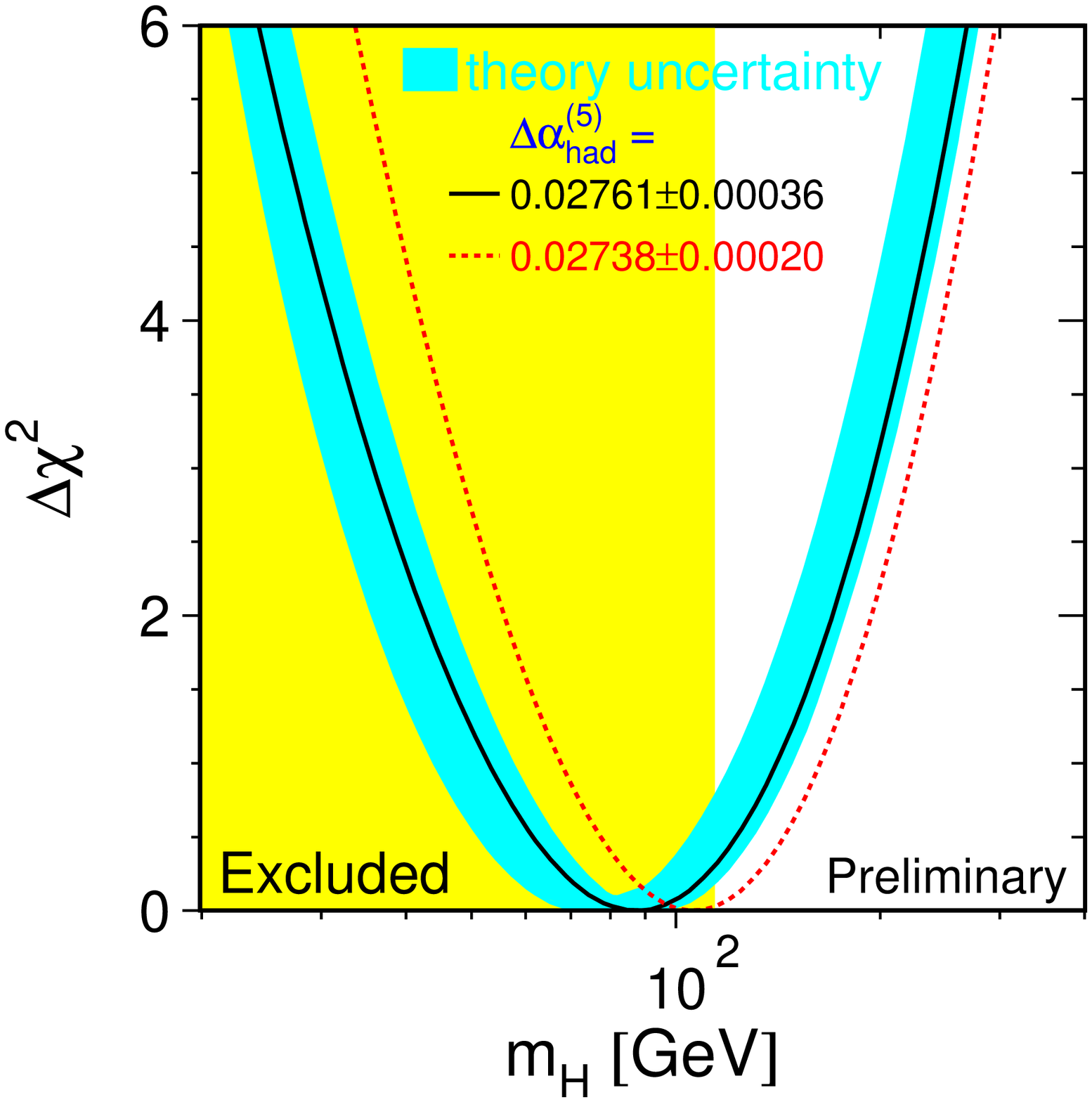}
      \includegraphics*[scale=0.35]{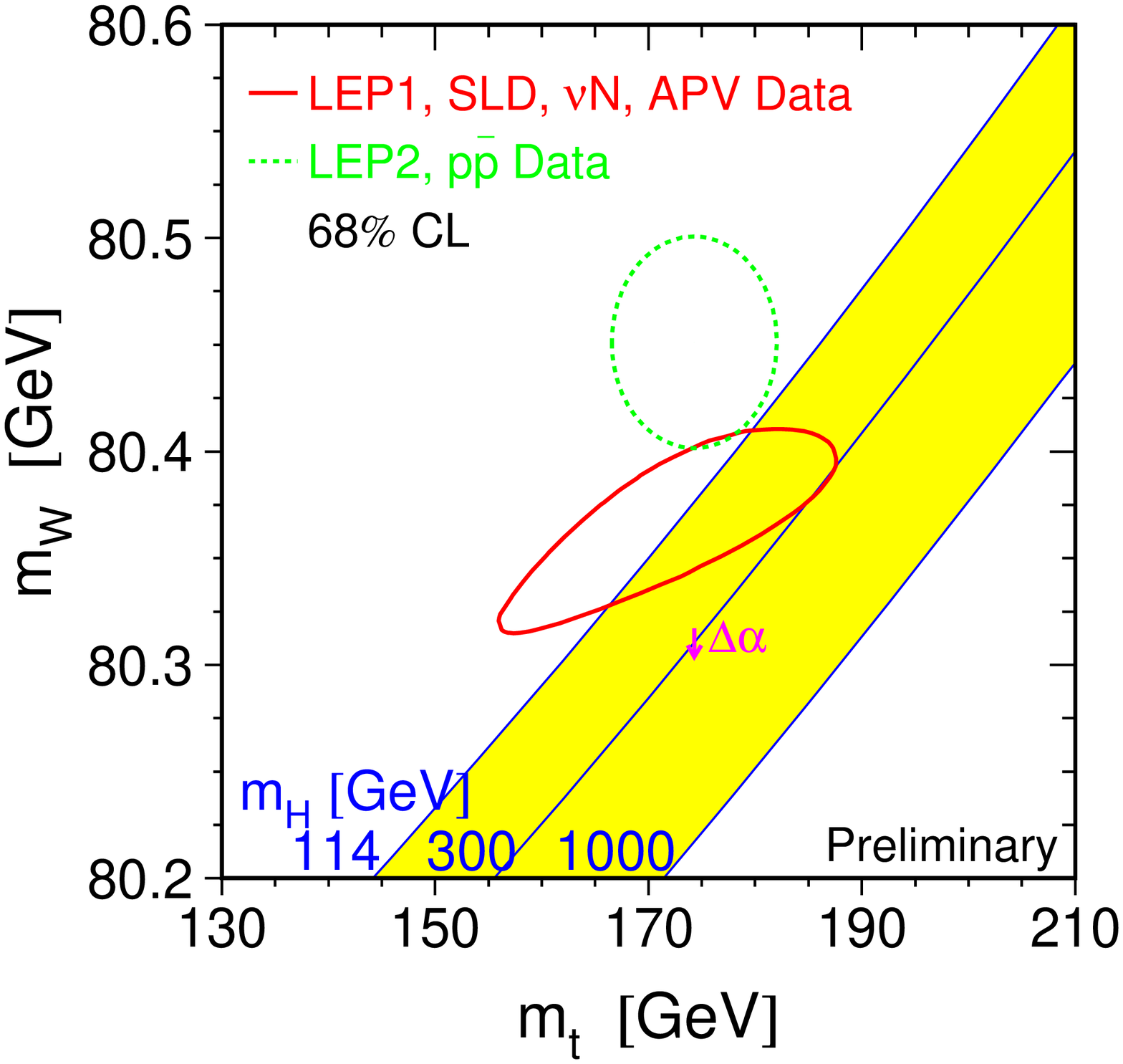}
  }
\caption{\em (left panel) $\Delta \chi^2$ as a function of $m_h$ 
for a fit to the SM of the precision measurements at LEP and (right panel) 
consistency between the direct and indirect measurements of 
$m_t,m_W$~\protect\cite{LEPC}, 
\label{fig4}}           
\end{figure}
listed in Fig.\ref{fig1} on $m_{h}$. 
The left panel of Fig.~\ref{fig4}~\cite{LEPC} shows the 
value of $\Delta \chi^{2}$ for the SM fits as a function of $m_{h}$. Boundary 
of the shaded area indicates the lower limit on $m_{h}$  implied by direct 
measurements at LEP2, $m_{h}$ $<$
113 GeV. Consistency of the upper limit of 196 GeV (222 GeV) 
for two different choices of $\Delta\alpha_{had}$  listed on 
the figure, with the range of 160
$\pm$ 20 geV predicted by the SM (cf. Fig. 3 left panel) for 
$\Lambda$ = $M_{pl}$ is very tantalizing. The two ovals in
the right panel of the Fig. 4 show the indirectly and directly
measured values of $m_t,m_W$ along with the lines of the SM predictions
for different Higgs masses. First this shows clearly that a light Higgs
is preferred strongly in the SM. Further we can see that an improvement in the
precision of the $m_W$  and $m_t$ measurement will certainly help give 
further indirect information on the Higgs sector and hence will allow probes of 
the physics beyond the SM, if any should be indicated  by the data in future 
experiments. Note that the $m_H$ in the labels on the axes in
Figs. \ref{fig3} and \ref{fig4} is the same as $m_h$ used in the text.

Recently more general theoretical analyses of correlations of the scale of
new physics and the mass of the Higgs have started\cite{murayama,barbie}. 
In these the assumption is that the SM is only an effective theory and 
additional higher dimensional operators can be added. Based on very general 
assumptions about the coefficients of these higher dimensional operators, 
an analysis of the precision data from LEP-I with their contribution to the
observables, 
alongwith a requirement that the radiative corrections to the $m_h$ do not
destabilize it more than a few percent, allows different regions in $m_h -
\Lambda$ plane. This is shown in the right panel in 
Fig.~\ref{fig3}.  The lesson to learn from this figure taken from 
Ref.~\cite{murayama} is  
that a light higgs with $m_h < 130$ GeV will imply existence of new 
physics at the scale $\Lambda < 2-3 $ TeV, whereas  $195 < m_h < 215$ GeV 
would imply $\Lambda_{Np} < 10$ TeV.

We thus see that the theoretical consistency of the SM as a field theory
with a light Higgs implies scale of new physics between 1 to 10 TeV.
Conversely, the Higgs mass itself is severely constrained if SM is the right
description of physics at the EW scale (at least as an effective theory)
which is the case as shown by the experiments. If the physics that tames the
bad high energy behaviour of the EW amplitudes is not given by the Higgs
scalar, it would imply that there exist some nonperturbative physics giving
rise to resonances corresponding to a strongly interacting $W^{+}W^{-}$ 
sector. This should happen at around a TeV scale. 
 
Such a sector is implied in a formulation of a gauge invariant
theory of  EW interactions, without the elementary Higgs. However, 
in this case, one has to take recourse to the nonlinear realization of 
the symmetry using the Goldstone Bosons, but of course one with custodial 
symmetry which will maintain $\frac{M^{2}_{W}}{M^{2}_{Z} cos^{2}~\theta_{W}}=1$.
One uses, (for notations and a mini review see \cite{Morioka})
\beqn
\Sigma=exp(\frac{i \omega^i \tau^i}{v}) \;\;\; {{\cal D}}_{\mu}
\Sigma=\partial_\mu \Sigma + \frac{i}{2} \left( g \W_{\mu} \Sigma
- g'B_\mu \Sigma \tau_3 \right)
\eeqn
The $W,Z$ masses are simply
\beqn
{\cal L}_M=\frac{v^2}{4} \tr({\cal d}^\mu \Sigma^\dagger {\cal d}_\mu \Sigma)
\eeqn
which is the lowest order operator one can write. \\

The consistency of such a model with the LEP data which seems to require a
light higgs is possible since the above mentioned fits are valid only within
the SM. The fits to the precision measurements in this case are analysed in
terms of the oblique parameters S, T, U~\cite{takeuchi}. It is necessary to 
consider additional new, higher dimensional operator which would give 
negative S~\cite{Fawzi,Morioka}
\beqn
{{\cal L}}_{10}&=&g g' \frac{L_{10}}{16 \pi^2} \tr ( \B^{\mu \nu}
\Sigma^{\dagger} \W^{\mu \nu}  \Sigma ) \longrightarrow
L_{10}=-\pi S_{\rm New}
\eeqn                                                             
which breaks down the custodial symmetry somewhat. The earlier discussion 
of the higher dimensional operators involving both gauge and fermionic fields
~\cite{barbie} is a case where these ideas are taken a bit further. Howver,
the fermionic operators were found to be strongly constrained there.
The additional bosonic operators  that one can write  
contribute to the trilinear and quartic couplings of the gauge
bosons though not to the $S,T$ and $U$. For 
instance, ${{\cal L}}_{9L}=-i g \frac{L_{9L}}{16 \pi^2} \tr (
\W^{\mu \nu}\cd_{\mu} \Sigma \cd_{\nu} \Sigma^{\dagger} )$ and
$\cl_{1}=\frac{L_1}{16 \pi^2} \left( \tr (D^\mu \Sigma^\dagger
D_\mu \Sigma) \right)^2$ to cite only two (for more see
\cite{Morioka}). Now these operators need to be probed at higher
energies. In order that one learns more than what we have with the
LEP data, these operators should be constrained better than
$L_{10}$, {\it i.e.}, the $L_i$ should be measured better than
$.1$, ideally one should aim at the $10^{-2}$ level. This is hard
since already $L_{9L}\sim .1$ implies measuring the $\Delta
\kappa_\gamma$ in the $WW\gamma$ vertex at $\Delta \kappa_\gamma
\sim 1.3 \;10^{-4}$
So if there is no elementary Higgs, the new physics will show up in
modification of the trilinear/quartic gauge boson couplings. The scale of
this new physics, allowed by the LEP data is again $\leq$ 30 TeV.

The above arguments indicate clearly that the demands of the theoretical
consistency of the SM and the excellent agreement of {\it all} the 
high precision
measurements in the EW and strong sector, imply some new physics at an
energy scale $\sim$ TeV, which should hold clues to the phenomenon of the
breaking of the EW symmetry. A TeV scale collider is thus necessary to
complete our understanding of the fundamental interactions.

Quantum field theories with scalars, like the higgs scalar $h$
(incidentally, $h$ is the only scalar in the SM) have problems of 
theoretical consistency, in the sense that the mass of the scalar 
$m_{h}$ is not stable under radiative corrections. Under the prejudice 
of a unification of all the fundamental interactions, one expects a 
unification scale $\sim$ $10^{15}$ - $10^{16}$ GeV. Even in the
absence of such unification, there exists at least one high scale in particle
theory, viz., the Plank scale ($M_{pl}$ = $10^{10}$ GeV) where gravity 
becomes strong.  Since the existence of $h$ is related to the phenomenon of 
EW symmetry breaking, $m_{h}$ is bounded by TeV scale as argued above. One 
needs to fine tune the parameters of the scalar potential order by order 
to stabilize $m_{h}$ at TeV scale against large contributions coming from 
loop corrections proportional to $M_{U}^{2}$ where $M_{U}$ is the high scale. 
This fine tuning can be avoided~\cite{romesh} if protected by Supersymmetry
(SUSY), a symmetry relating bosons with fermions. 
In this case the scalar potential is completely determined in terms of the 
gauge couplings and gauge boson masses. It can be written as 
\beqn
V = {|\mu|}^{2} \left(  {|H_1|}^{2} + {|H_2|}^{2}
 \right) + \frac{g^2+g^{'2}}{8} {\left( {|H_1|}^{2} -
  |H_2|^2 \right)}^{2} + \frac{g^2}{2} {|H_1^*H_2|}^{2} \geq 0\nonumber
\eeqn
Note the appearance of the $\mu$ term which is a SUSY conserving
{\em free} parameter. But note also that the quartic couplings are
gauge coupling. So one must add (soft) SUSY breaking parameters in
such a way that one triggers electroweak symmetry breaking.
\bea V_H
&=& (m_{11}^2|+\mu|^2 ) |H_1|^2 + (m_{22}^2+|\mu|^2)  |H_2|^2 -
m_{12}^2 \epsilon_{ij} \left( H_1^i H_2^j + h.c. \right) \nonumber
\\ &+&
\frac{g^2+g^{'2}}{8} \left( |H_1|^2 - |H_2|^2 \right)^2 \;+\;
\frac{g^2}{2} |H_1^* H_2|^2
\eea                                                             
  At tree level, this means $m_{h}$ $<$  $m_{Z}$. Loop corrections due to large
$m_{t}$ as
well as due to Supersymmetry breaking terms, push this upper limit to 
$\sim$ 135 GeV in the Minimal Supersymmetric Standard Model 
(MSSM)~\cite{heinemeyer,quiros}.
Thus if supersymmetry exists we certainly expect the higgs to be within the
reach of the TeV colliders.  As a matter of fact the  analysis
discussed earlier~\cite{murayama} relating the scale of new physics and higgs 
mass, specialized to the case of Supersymmetry, do imply that Supersymmetry 
aught to be at TeV scale to be relevant to solve the fine tuning problem. 
If Supersymmetry has to provide
stabilization of the scalar higgs masses in a natural way~\cite{barbieri} 
some part of the sparticle spectra has to lie within TeV range. Even in 
the case of focus point supersymmetry with superheavy scalars ~\cite{matchev} 
the charginos/neutralinos and some of the sleptons are expected to be 
within the TeV Scale.

In the early days of SUSY models there existed essentially only one 
one class of models where the  supersymmetry breaking is transmitted 
via gravity to the
low energy world. In the past few years there has been tremendous 
progress in the ideas about SUSY breaking and thus there exist now a 
set of different models and the different physics they embed is reflected
in difference in the expected structure and values  of the soft supersymmetry 
breaking terms. Some of the major parameters are the masses of the extra
scalars and fermions  in the theory  whose masses are generically 
represented by $M_0$ and $M_{1/2}$. 

\bedes
\item[1)] Gravity mediated models like minimal supersymmetric extension
of the standard model (MSSM),  supergravity model (SUGRA) where the 
supersymmetry breaking is induced  radiatively etc.
Both assume universality of the gaugino and sfermion 
masses  at the high scale.  In this  case supergravity couplings 
of the fields in the hidden sector with the SM fields are responsible for 
the soft supersymmetry breaking  terms.  These models always have  
extra scalar mass parameter $m_0^2$ which  needs fine tuning so that 
the sparticle exchange does not generate the unwanted flavour changing neutral 
current (FCNC) effects, at an unacceptable level.  
\item[2)]In the Anomaly Mediated Supersymmetry Breaking (AMSB) models 
supergravity couplings which cause mediation are absent and the 
supersymmetry breaking  is caused by loop effects.  The conformal
anomaly generates the soft supersymmetry breaking and the sparticles 
acquire masses due to the breaking of scale invariance. 
This mechanism becomes a viable one  for solely generating the supersymmetry
breaking  terms, when the quantum contributions to the gaugino masses
due to the `superconformal anomaly' can be large~\cite{RS2,GR}, hence
the name Anomaly mediation for them.  The slepton masses in this model are 
tachyonic in the absence of a scalar mass parameter $M_0^2$.
\item[3)] An alternative scenario where the soft supersymmetry breaking 
 is transmitted to the low
energy world via a  messenger sector   through  messenger fields which have 
gauge interactions, is called the Gauge Mediated Supersymmetry Breaking 
(GMSB)~\cite{gmsb_rev}. These models have no problems with the FCNC 
and do not involve any scalar mass parameter.
\item[4)]  There exist also a  class of models where the mediation of the
symmetry breaking is dominated by gauginos~\cite{gaumsb}. 
In these models  the matter sector feels the effects of SUSY breaking
dominantly via gauge superfields. As a result, in these scenarios, one expects 
$M_0 \ll M_{1/2}$, reminiscent of the `no scale' models.

\edes
All these models clearly differ in their specific predictions for various 
sparticle spectra, features of some of which are summarised in 
\begin{table}[htb]
\caption{The table gives predictions of different types of SUSY breaking
models for gravitino, gaugino and scalar masses $\alpha_{i} =
{g_{i}^{2}}/{4 \pi}$ (i=1,2,3 corresponds to U(1), SU(2) and SU(3)
respectively), $b_{i}$ are the coefficients of the
${-g_{i}^{2}}/{(4 \pi )^{2}}$ in the expansion of the $\beta$ functions 
$\beta_{i}$ for  the coupling $g_i$ and $a_i$ are the coeffecients 
of the corresponding expansion of the anomalous dimension. the coeffecients 
$D_i$ are the squared gauge charges multiplied by various factors which 
depend on the loop contributions to the scalar masses in the different models.}
\vspace{0.2cm}
\begin{tabular}{cccc}
&&&\\
\hline
&&&\\
 Model & $M_{\tilde{G}}$ & $(mass)^2$ for gauginos & $(mass)^2$ for scalars\\
&&&\\
\hline
&&&\\
 mSUGRA & ${{M_{S}^{2}} / {\sqrt{3} M_{pl}}}$ $\sim $ TeV &
$({\alpha_{i}}/{\alpha_{2}})^{2}$ $M_{2}^{2}$ & $M_{0}^{2} + \sum_{i}
D_{i} M_{i}^{2}$ \\
cMSSM & $M_{S} \sim 10^{10} - 10^{11}$ GeV & $\mbox{ }$ & $\mbox{  }$ 
\\
&&&\\
\hline
&&&\\
GMSB & $({\sqrt{F}}/{100TeV})^{2}$ eV &
$({\alpha_{i}}/{\alpha_{2}})^{2} M_2^2$ & $\sum_{i} D_{i}^{'} M_{2}^{2}$ \\
$\mbox { }$  & 10 $<  \sqrt{F} < 10^4 $ TeV &  & \\ 
&&&\\
\hline
&&&\\
AMSB & $\sim $ 100 TeV & 
$({\alpha_{i}}/{\alpha_{2}})^{2} ({b_{i}}/{b_{2}})^{2} M_2^2$ & 
$\sum_{i} 2 a_{i} b{i} ({\alpha_{i}}/{\alpha_{2}})^{2} M_2^2$ \\ 
&&&\\
\hline
\end{tabular}
\label{T:rgplen:1}
\end{table}
Table~\ref{T:rgplen:1} following ~\cite{peskin_talk}, 
where $M_1,M_2$ and $M_3$ denote masses of the fermionic partners 
of the $U(1),SU(2)$ and $SU(3)$ gauge bosons respectively
and the  messenger scale parameter $\Lambda$ more generally used in GMSB models 
has been traded for $M_2$ for ease of comparison  among the different models.
As one can see the expected mass of the gravitino, the 
supersymmetric  partner of the spin 2 graviton, 
varies widely in different models. The SUSY breaking scale
$\sqrt{F}$ in GMSB model is restricted to the range shown in Table 1 by
cosmological considerations. Since $SU(2), U(1)$ gauge groups are not
asymptotically free,\ie, $b_i$ are negative, the slepton masses are tachyonic
in the AMSB model, without a scalar mass parameter, as can be seen from 
the third column of the table. The minimal cure to this is, as mentioned 
before, to add an additional parameter $M_{0}^{2}$, not shown in the table, 
which however spoils the invariance of the mass relations between various
gauginos (supersymmetric partners of the gauge bososns) under the  different 
renormalisations that the different gauge couplings receive. In  the 
gravity mediated models like mSUGRA, cMSSM and most of the versions of 
GMSB models, there exists gaugino mass unification at high scale, whereas 
in the AMSB models the gaugino masses are given by Renormalisation Group
invariant equations and hence are determined completely by the values of the 
couplings at low energies and become ultraviolet insensitive. Due to this 
very different scale dependence,
the ratio of gaugino mass parameters at the weak scale in the two sets of
models are quite different: models 1 and 2 have $M_1 : M_2 : M_3$ = 1 : 2 :
7 whereas in the AMSB model one has $M_1 : M_2 : M_3$ = 2.8 : 1 : 8.3.
The latter therefore, has the striking prediction that the lightest chargino
$\Chipm_{1}$ (the spin half partner of the $W^\pm$ and the charged higgs 
bosons in the theory) and the lightest supersymmetric particle (LSP) $\N0_{1}$, are almost  pure
SU(2) gauginos and are very close in mass. The expected particle spectra in
any given model can vary a lot. But still one can make certain general
statements,\eg the ratio of squark masses to slepton masses is usually 
larger in the GMSB models as compared to mSUGRA.
In mSUGRA one expects the sleptons to be lighter than the first two
generation squarks, the LSP is expected mostly to be an (U1) gaugino
and the right
handed sleptons are lighter than the left handed sleptons. On the other
hand, in the AMSB models, the left and right handed sleptons are almost
degenerate. Since the crucial differences in different models exist in 
the slepton
and the chargino/neutralino sector,  it is clear that the leptonic 
colliders which can study these sparticles with the EW interactions, with
great precision, can  play really a crucial role in being able to distinguish
among different models.

The above discussion, which illustrates the wide `range' of predictions of
the SUSY models,  also makes it clear that  a general discussion of the  
sparticle phenonenology  at any collider is far too complicated. To me, 
that essentially reflects our ignorance. This makes it even more imperative 
that we try to extract as much model independent information from the 
experimental measurements. This is one aspect where the leptonic 
colliders can really play an extremely important role.

 The recent theoretical developments in the subject of `warped large' extra
dimensions~\cite{RS} provide a very attractive solution to the
abovementioned hierarchy problem by obviating it as in this case gravity
becomes strong at TeV scale. Hence we do not have any new physics as from
Supersymmetry, but then there will be modification of various SM couplings
due to the effects of the TeV scale gravity. Similarly, depending on the
particular formulation of the theory of `extra' dimensions~\cite{RS,ADD},  
one expects to have new particles in the spectrum with TeV scale masses as 
well as with spins higher than 1.   These will  manifest themselves as 
interesting phenomenology at the future colliders in the form of additional
new, spin 2 resonances or modification of four fermion interaction or
production of high energy photons etc.

\section { Search for the Higgs }
\subsection*{ Search for the SM higgs at Hadronic Colliders}
\begin{figure}[htb]                
       \centerline{
      \includegraphics*[scale=0.50]{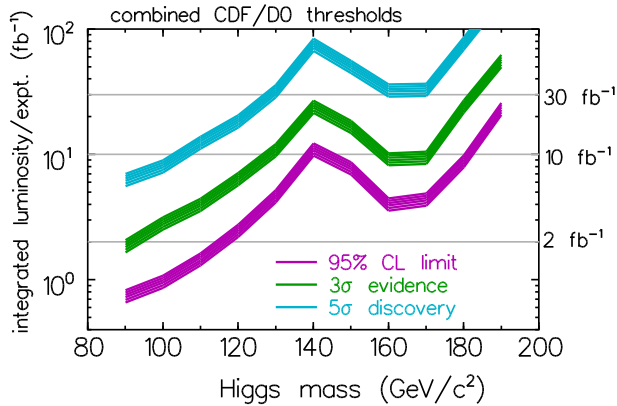}
  }
\caption{Tevatron Expected discovery/exclusion mass limit on the 
Higgs mass at the Tevatron\cite{TevatronHiggsWG2000})
\label{fig5}}           
\end{figure}

The current limit on $m_h$ from precision measurements at LEP is 
$m_h < 210$ GeV at $95\%$ C.L. and limit from direct searches is 
$m_h\lesssim 113$ GeV\cite{LEPC}. Tevatron is likely to be able to 
give indications  of the existence of a SM higgs,  by combining data in  
different channels together for $m_h \lesssim 120$ GeV if 
Tevatron run II can accumulate $30  fb^{-1}$ by 2005. This is shown in
Fig.~\ref{fig5}\cite{TevatronHiggsWG2000}

In view of the discussions of the expected range for $m_h$ of the 
last section as well as the current LEP limits on its mass, LHC is really
the collider to search for the Higgs where as Tevatron might just see 
some indication for it. 
The best mode for the detection of Higgs depends really on its mass. Due to
the large value of $m_t$ and the large $gg$ flux at LHC, the highest production
cross-section is via $gg$ fusion. 
\begin{figure}[htb]
\begin{center}
\includegraphics*[scale=0.3]{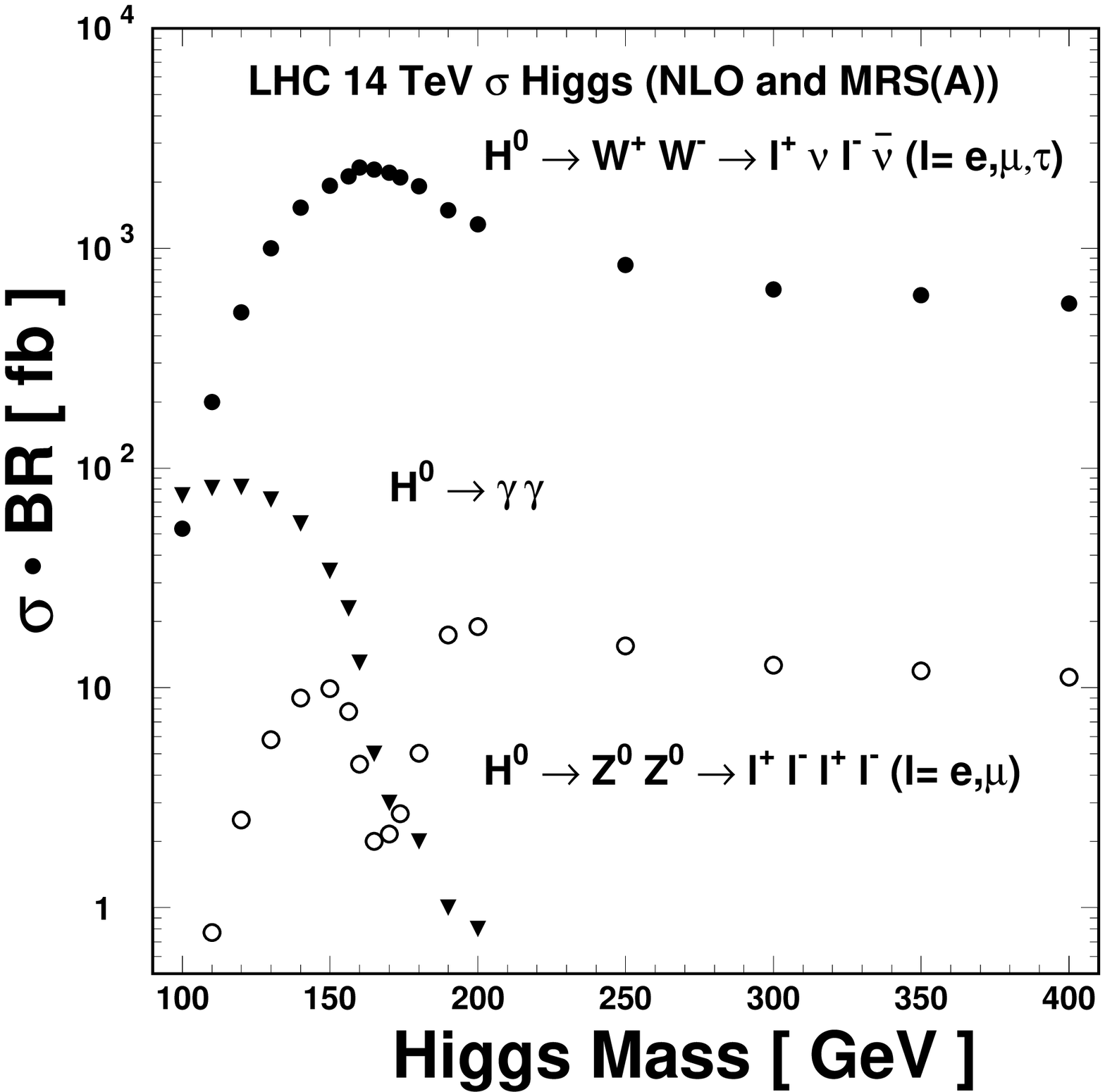}
\caption{\em Expected $\sigma \cdot BR$ for different detectable SM Higgs
decay modes~\protect\cite{michael}.}
\label{fone}
\end{center}
\end{figure}
Fig.~\ref{fone}\cite{michael} shows 
$\sigma \cdot BR$ for the SM higgs for various final states. 
The  search prospects are optimised by exploring different channels
in different mass regions.  The inclusive channel using $\gamma \gamma$ 
final states corresponds to  $\sigma \cdot  BR \; {\rm of\;\; only} \:  50 \: fb$,
but due to the low background it constitutes the cleanest channel for 
$m_h < 150$ GeV.
The important detector requirement for this measurement is good resolution
for $\gamma \gamma$ invariant mass. The detector ATLAS  at LHC 
should be able to achieve $\sim 1.3$ GeV whereas the detector  CMS 
expects to get $\sim 0.7$ GeV.  A much more interesting channel is 
production of a higgs recoiling against a jet.
The signal is much lower but is also much cleaner. 
Use of this channel gives a significance of $\sim 5 \sigma$
already at $30  fb^{-1}$ for the mass range $110 < m_h < 135 $ GeV.
A more detailed study of the channel $  pp \to h + t \bar t \to \gamma \gamma +
t \bar t$ is important also for the measurement of the $h t \bar t$ couplings.
For larger masses $(m_h \gsim 130$  GeV ) the channel 
$gg \to h \to Z Z^{(*)} \to 4l$ is the best channel.  
Fig.~\ref{fone} shows that, in the range $150 \: {\rm GeV}  < m_h < 190 $ 
GeV this clean channel, however, has a rather low $(\sigma \cdot  BR)$.  
The viability of $p \bar p  \to W W^{(*)} \to l \bar \nu \bar l \nu$
in this range has  been demonstrated\cite{michael}.
Thus to summarise for $m_h \lsim 180 $ GeV, there exist 
a large number of complementary channels whereas beyond that the 
gold plated  $4 l$ channel is the obvious choice. If the Higgs is
heavier, the  event rate will be too small in this channel 
({\it cf.} Fig.~\ref{fone}). Then the best option is to tag the forward jets
by studying the production of the Higgs in the process $ p \bar p \to W W q \bar q \to q \bar q h$.
\begin{figure}[htb]                
       \centerline{
             \includegraphics*[scale=0.38]{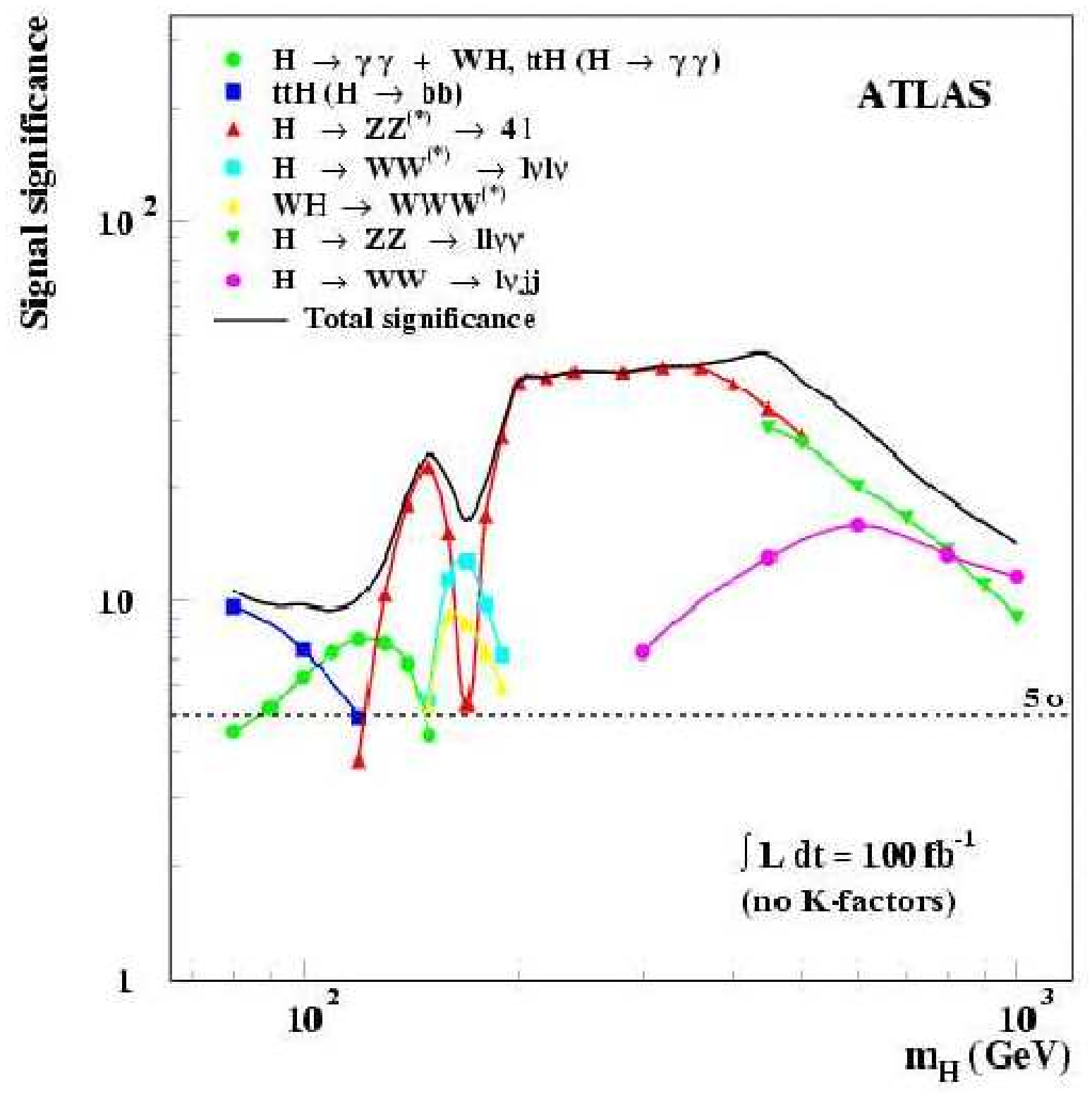}
              \includegraphics*[scale=0.45]{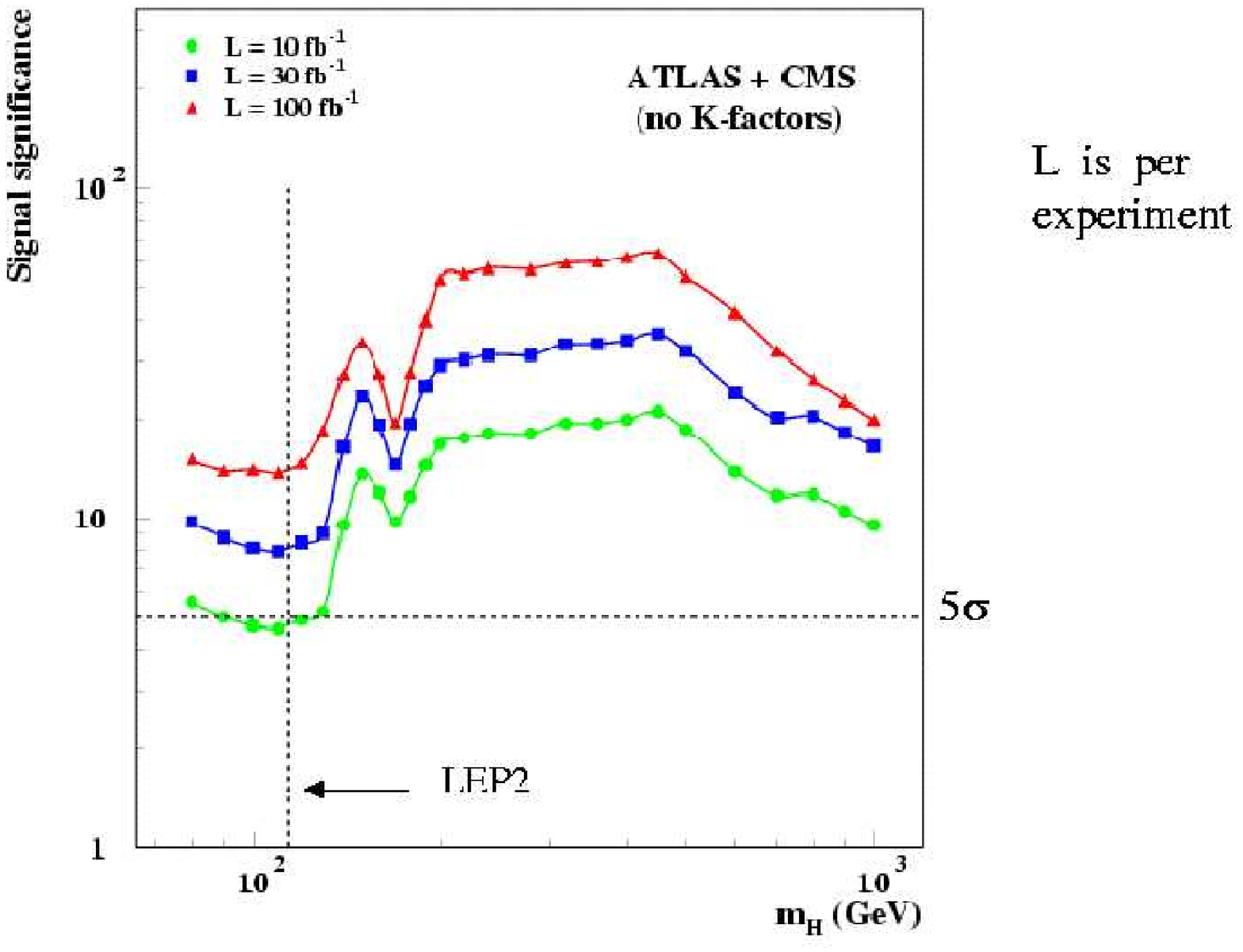}
                   }
        \caption{\em The expected significance level of the SM Higgs
signal at LHC~\protect\cite{fabiola}.\label{ftwo}}           
\end{figure}

The figure in the left panel in fig.~\ref{ftwo} shows the overall discovery 
potential of the SM
higgs in all these various channels whereas the one on the right shows the same
overall profile of the significance for discovery of the SM higgs, for three
different luminosities, combining the data that both the detectors ATLAS and CMS
will be able to obtain. The figure shows that the SM higgs boson can be 
discovered
(i.e. signal significance $\gsim 5$) after about one year of operation 
even if $m_h \lsim 150$ GeV. Also at the end of the year the SM higgs boson 
can be ruled out over the entire mass range implied in the SM discussed earlier.

A combined study by CMS and ATLAS shows that a measurement of $m_h$ at $0.1\%$
level is possible for $m_h \lsim 500$ GeV, at the end of three years of high 
luminosity run and is shown in
\begin{figure}[htb]
        \centerline{
             \includegraphics*[scale=0.40]{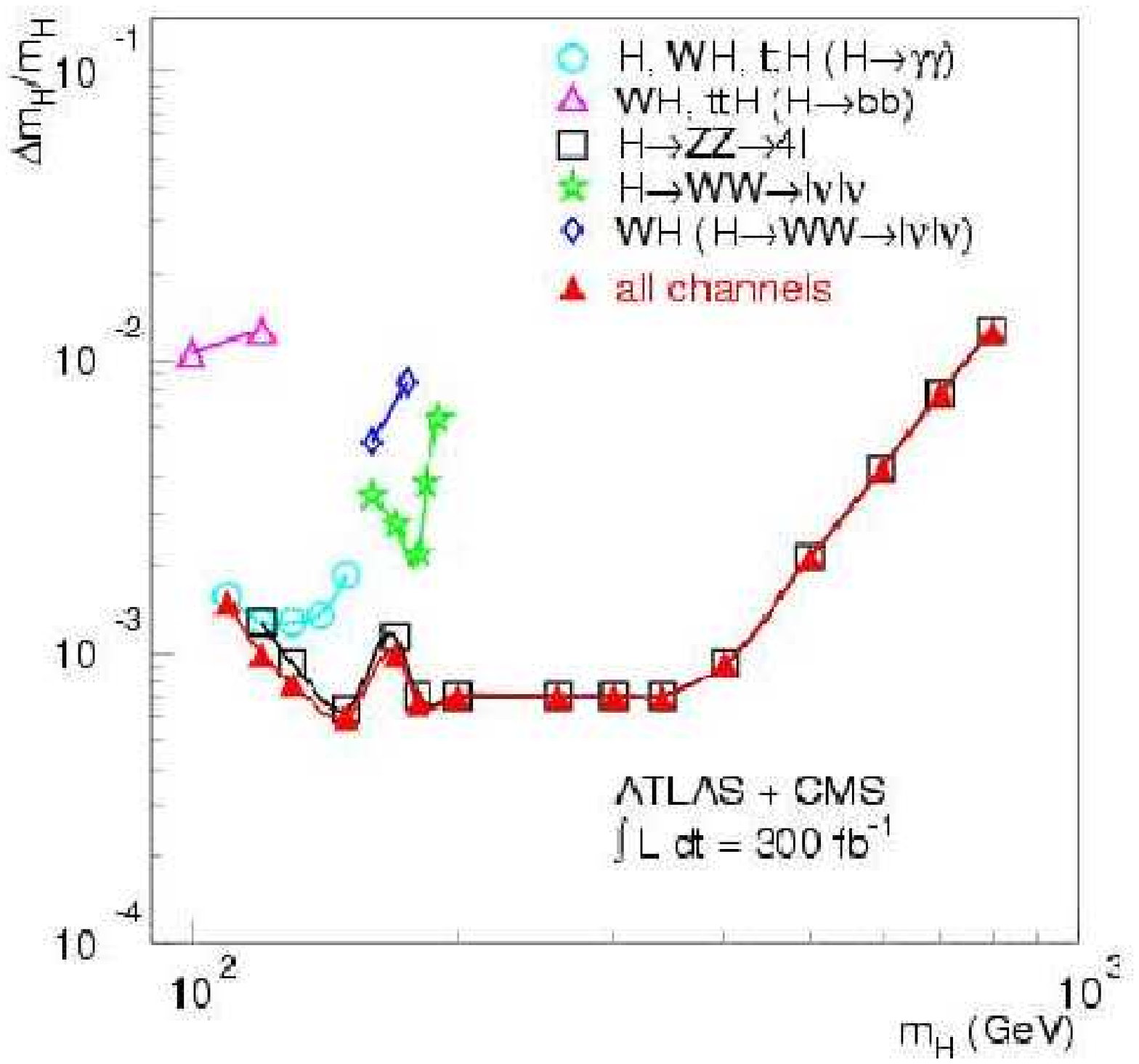}
              \includegraphics*[scale=0.38]{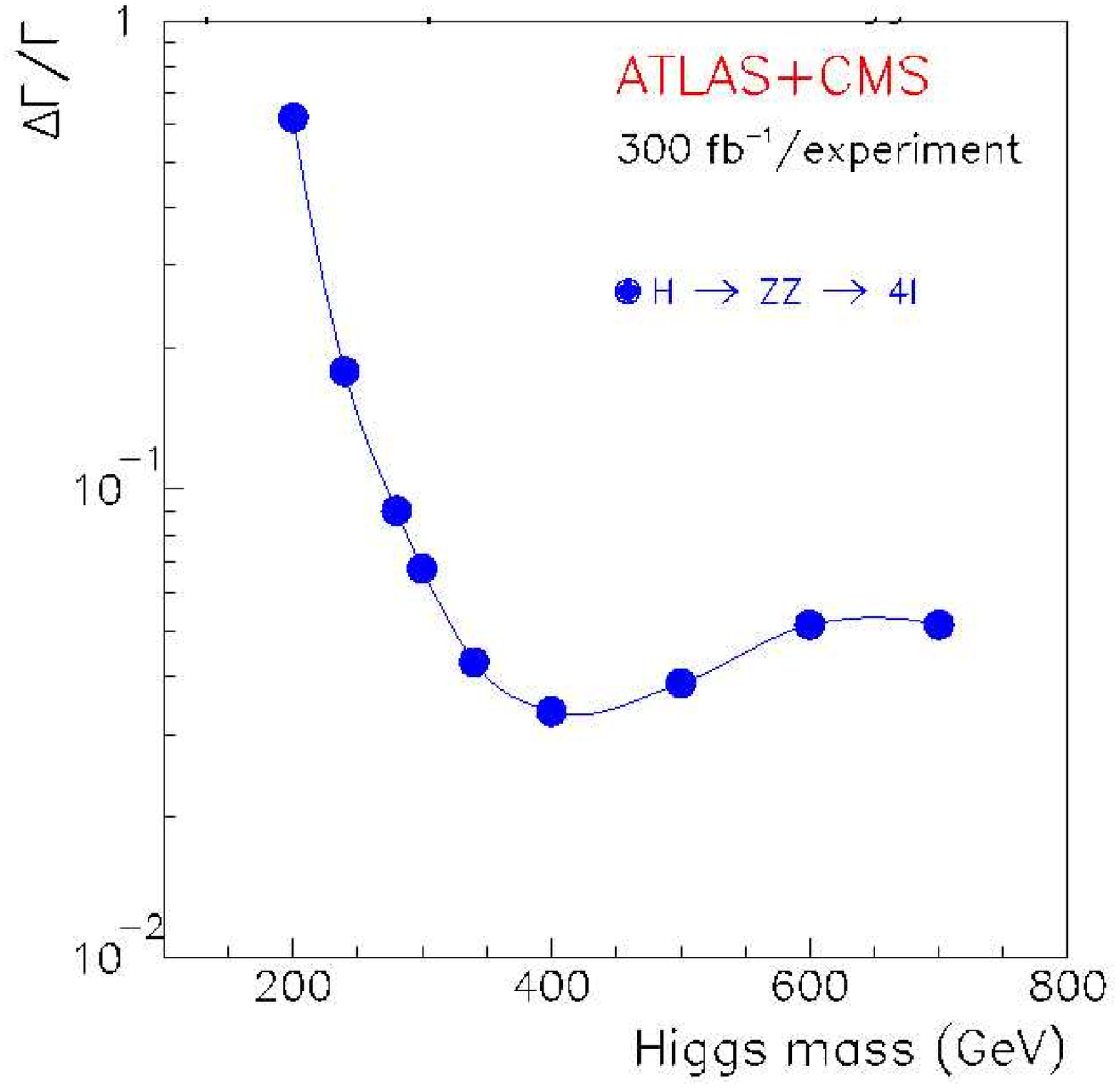}
}
\caption{\em Expected  accuracy of the measurement of Higgs mass and 
width at LHC\protect\cite{fabiola}.\label{fthree}}
\end{figure}
fig.~\ref{fthree}. As far as the width $\Gamma_h$ is concerned, a measurement 
is possible only for $m_h > 200$ GeV, at a level of $\sim 5\%$ and that too
at the end of three years of the high luminosity run. 
The values of
$\Delta \Gamma / \Gamma$ that can be reached at the end of three years of high
luminosity run, obtained in a combined ATLAS and CMS study, are shown in 
the figure in the right panel in fig.~\ref{fthree}. 

Apart from the precision measurements of the mass and the width of the 
Higgs particle, possible accuracy of extraction of the couplings of the
Higgs with the matter and gauge particles, with a view to  check the 
spontaneous symmetry breaking scenario, is also an important issue. 
\begin{table}[htb]
\caption{\em Expected accuracy in the extraction of the Higgs couplings as
evaluated by ATLAS\protect\cite{fabiola}.\label{tab:tab2}}
\begin{center}
\begin{tabular}{|c|c|c|}
\hline
Ratio of cross-sections&Ratio of extracted &Expected Accuracy\\
&Couplings&Mass Range\\
\hline
&&\\
$\frac{\sigma (t \bar t h + Wh)(h \to \gamma \gamma)}{\sigma (t \bar t h + Wh)(h \to b \bar b)}$&$\frac{B.R.(h \to \gamma \gamma)}{B.R.(h \to b  \bar b)}$&$\sim 15\%$, 80-120 GeV\\
&&\\
\hline
&&\\
$\frac{\sigma (h \to \gamma \gamma)}{\sigma (h \to 4 l)}$&
$\frac{B.R.(h \to \gamma \gamma)}{B.R.(h \to Z Z^{(*)})}$&$\sim 7\%$, 
120-150 GeV\\
&&\\
\hline
&&\\
$\frac{\sigma (t \bar t h \to \gamma \gamma/b \bar b)}
{\sigma ( Wh \to \gamma \gamma /b \bar b)}$& $\frac{g^2_{h t \bar t}}{g^2_{hWW}}$&$\sim 15\%, 80 < m_h < 120$ GeV\\
&&\\
\hline
&&\\
$\frac{\sigma (h \to Z Z^*\to 4 l)}{\sigma ( h \to W W^* \to l \nu l \nu)}$& 
$\frac{g^2_{hZZ}}{g^2_{hWW}}$&$\sim 10\%, 130 < m_h < 190$ GeV\\
&&\\
\hline
\end{tabular}
\end{center}
\end{table}
Table~\ref{tab:tab2} shows the accuracy which would be possible in 
extracting ratios of various couplings, according to an analysis by 
the ATLAS collaboration. This analysis is done by measuring the 
ratios of cross-sections so that the measurement is insensitive to 
the theoretical uncertainties in the prediction of hadronic 
cross-sections. All these measurements use only the inlclusive 
Higgs mode. 

New analyses based on an idea by Zeppenfeld and 
collaborators\cite{dieter-mad} have explored the
use of production of the Higgs via WW/ZZ(IVB) fusion, in the process
$pp \to q + q + V + V + X \to q + q + h + X.$ Here the two jets go in the 
forward direction.
This has increased the possibility of studying the Higgs production via 
IVB fusion process to lower values
of $m_h (< 120 {\rm GeV})$ than previously thought possible. 
It has been demonstrated\cite{dieter-mad} that using the 
production  of Higgs in the process $qq \to h jj$, followed by the decay
of the Higgs into various channels $\gamma \gamma, \tau^+ \tau^-, W^+ W^-$ 
as well as the inclusive channels $gg \to h \gamma \gamma, gg \to h 
\to Z Z^{(*)}$, it should be possible to measure $\Gamma_h, g_{hff} $ 
and $g_{hWW}$ to a level of $10-20 \%$ , assuming
that $\Gamma (h \to b \bar b) / \Gamma (h \to \tau^+ \tau^-)$ has approximately
the SM value. Recall, here that after a full LHC run,with a combined CMS +ATLAS
analysis, the latter should be known to $\sim 15 \%$. 
In principle, such measurements of the Higgs couplings might be an indirect way
to look for the effect of physics beyond the SM. We will discuss this later.

By the start of the LHC with the possible TeV 33 run with
$\int {\cal L} dt = 30  fb^{-1}$, Tevatron can give us an indication and 
a possible signal for a light Higgs, combining the information from 
different  associated production  modes: $Wh,Zh $ and $W W^{(*)}$. The 
inclusive channel $\gamma \gamma /4l$  which will be dominantly used at LHC is 
completely useless at Tevatron. So in  some sense the information we get from 
Tevtaron/LHC will be complementary. 

Thus in summary the LHC, after one year of operation should be able to see 
the SM higgs if it is in the mass range where the SM says it should be. Further
at the end of $\sim 6$ years the ratio of various couplings of $h$ will be 
known within $\sim 10 \%$.

\subsection*{Search for the SM higgs at \eplem\ colliders}
As discussed above LHC will certainly be able to discover the SM higgs
should it exist and study its properties in some detail as shown above. 
It is clear, however, that one looks to the clean environment of a $e^+e^-$ 
collider for establishing that the Higgs particle has all the properties 
predicted by the SM: such as its spin, parity, its couplings to the
gauge bosons and the fermions as well as the self coupling. Needless to 
say that this has been the focus of
the discussions of the physics potential of the future linear colliders
\cite{tesla,nlc,jlc}.  Eventhough we are not sure at present 
whether such colliders will become a reality,  the technical 
feasibility of buliding a $500$ GeV \eplem\ (and perhaps an  attendant
\gamgam, $e^-e^-$ collider) and doing physics with it is now 
demonstrated ~\cite{tesla,nlc,jlc}.
At these colliders, the production processes are $\eplem \to Z(^*) h
\to \ell^+ \ell^-  h$,  called Higgstrahlung,
$\eplem \to \nu \bar \nu h$  called $WW$ fusion and $\eplem \to t \bar t h$.
The associated production of $h$ with a pair of stops $\tilde t_1 \tilde t_1 h$
also has substantial cross-sections.  Detection of the Higgs at these machines 
is very simple if the production is kinematically allowed, 
as the discovery will be signalled by some very striking features of the 
kinematic distributions.  Determination of the spin of the  produced 
particle in this case  will also  be simple as the expected angular
distributions will be very different for scalars with even and odd parity. 
\begin{figure}[htb]
        \centerline{
             \includegraphics*[scale=0.40]{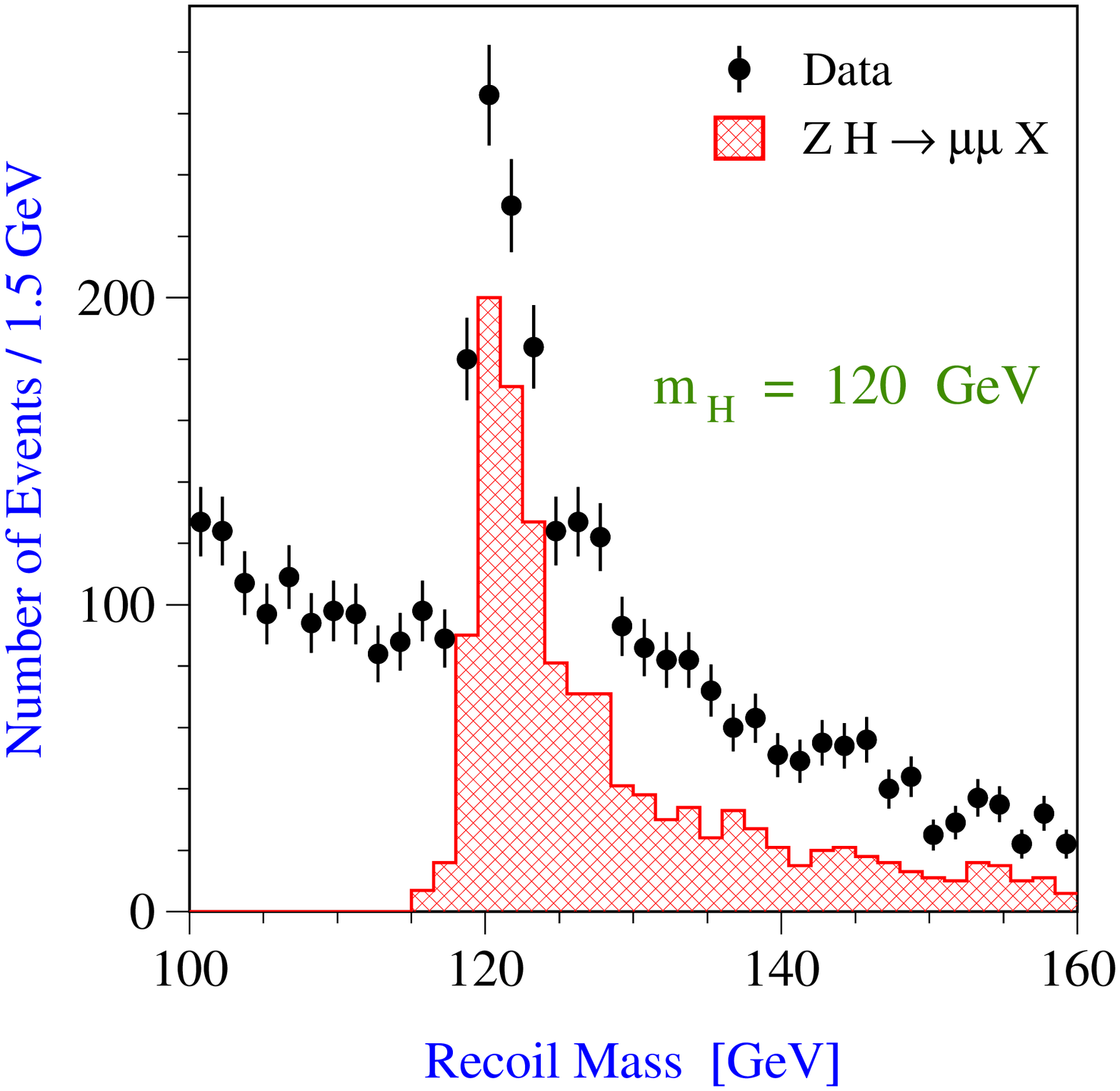}
}
\caption{The $\mu^+\mu^-$ recoil mass distribution in the process
$e^+ e^- \rightarrow h\ Z \rightarrow X \mu^+\mu^-$ for
$m_h$ = 120 GeV and 500 fb$^{-1}$ at $\sqrt{s}$ = 350 GeV. The dots with
error bars are Monte Carlo simulation of Higgs signal and background.
The shaded histogram represents the signal only~\protect\cite{tesla}}.
\label{teslah}
\end{figure}

At $\sqrt{s} = 350$ GeV, a sample of $\sim$ 80,000 Higgs bosons
is produced, predominantly through Higgs-strahlung,
for $m_h = 120$ GeV  with an integrated luminosity
of 500 fb$^{-1}$, corresponding to one to two years of
running. The Higgs-strahlung process, $e^+e^-\to Z h $,
with $Z \rightarrow \ell^+ \ell^-$, offers a very distinctive
signature~(see Fig.~\ref{teslah}) ensuring the observation of the SM Higgs
boson up to the production kinematical limit independently of its decay
(see Table~\ref{tab:discovery}).
At $\sqrt{s} = 500$ GeV, the Higgs-strahlung and the
$WW$ fusion processes have approximately the same cross--sections,
${\cal O}$(50 fb) for 100 GeV  $\lsim m_h \lsim$ 200 GeV.
\begin{table}[ht!]
\caption{Expected number of signal events for 500\,fb$^{-1}$ for the
Higgs-strahlung channel with di-lepton final states $e^+e^- \rightarrow
Z h \rightarrow \ell^+ \ell^- X$, ($\ell = e,~\mu$) at different $\sqrt{s}$
values and maximum value of $m_h$ yielding more than 50 signal events in this
final state.}
\begin{center}
\begin{tabular}{|c|c|c|c|}
\hline
$m_h$ (GeV) & $\sqrt{s}$ = 350 GeV & 500 GeV & 800 GeV \\
\hline
\hline
120     & 4670 & 2020 & ~740 \\
140     & 4120 & 1910 & ~707 \\
160     & 3560 & 1780 & ~685 \\
180     & 2960 & 1650 & ~667 \\
200     & 2320 & 1500 & ~645 \\
250     & ~230 & 1110 & ~575 \\ \hline
Max $m_h$ (GeV)  & 258 & 407 & 639 \\
\hline
\end{tabular}
\label{tab:discovery}
\end{center}
\end{table}
The very accurate measurements of quantum numbers of the Higgs that will be 
possible at such colliders can help distinguish between the SM higgs and the
lightest higgs scalar expected in the supersymmetric models. We will discuss 
that in the next section.

\subsection*{Search for the MSSM higgs at hadronic colliders}
The MSSM Higgs sector is much richer and has five scalars;
three neutrals: ${\cal CP}$ even \sh\ , H and ${\cal CP}$ odd A as well as 
the pair of charged
Higgses $H^\pm$. So many more search channels are available. The most important 
aspect of the MSSM higgs, however is the upper limit\cite{heinemeyer,quiros}
 of $130$ GeV ($200$ GeV) for MSSM (and its extensions), on
the mass of lightest higgs. The masses and couplings of these scalars depend
on the supersymmetric parameters  $m_A$, the mass of the ${\cal CP}$ odd Higgs
scalar $A$, the ratio of the vacuum expectation value of the two higgs 
fields $\tan \beta$ as well as SUSY breaking parameters $m_{\tilde t_1}$ and 
the mixing in the stop sector controlled
essentially by $A_t$. In general the couplings of the \sh\ can be quite 
different from the SM higgs $h$; {\it e.g.} even for large 
$m_A ( > 400 {\rm GeV}), \Gamma_{\sh} /\Gamma_{h} > 0.8,$ over most
of the range of all the other parameters. Thus such measurements can be 
a `harbinger' of SUSY. The upper limit on the mass of \sh\ forbids its
decays into a $VV$ pair and thus it  is much narrower than the SM $h$. Hence
the only decays that can be employed for the search of \sh\ are $b \bar b,
\gamma \gamma$ and $\tau^+ \tau^- $. 
The $\gamma \gamma$ mode can be suppressed  for the lightest supersymmetric
scalar \sh\ as compared to that to $h$ in the SM.  
The reduction is substantial even when all the sparticles are heavy,
at low $m_A, \tan \beta$. 
\begin{figure}[htb]
\begin{center}
\includegraphics*[scale=0.40]{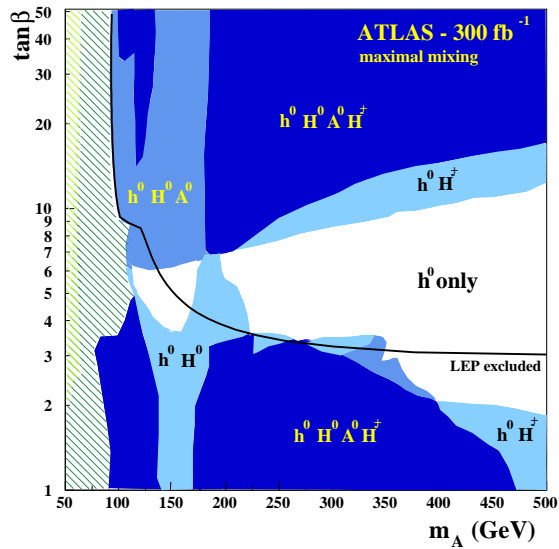}
\caption{Number of MSSM scalars observable at LHC  in different regions of
$\tan \beta - m_A$ plane\protect\cite{tesla}.}
\label{ffour}
\end{center}
\end{figure}
Fig.~\ref{ffour}, taken from Ref. \cite{tesla}, but which is essentially a 
rerendering  of a similar figure in Ref. \cite{fabiola}, shows  various regions 
in the $\tan \beta - m_A$ plane divided according to the number of the MSSM 
scalars observable at LHC, 
according to an ATLAS analysis, for the case of maximal mixing in the stop 
sector, at the end of full LHC run. This shows that for high $m_A (\gsim
200$ GeV) and low $\tan \beta (\lessapprox 8-9)$, only one of the five MSSM 
scalars will be observable.  Furthermore, the differences in the 
coupling of the SM and MSSM higgs are quite small in this region. Hence, it is 
clear that there exists part of the $\tan \beta - m_A$ plane, where the LHC
will not be able to see the extended Higgs sector of Supersymmetric models,
even though low scale Supersymmetry might be realised.

Situation can be considerably worse if some of the sparticles, particularly
$\tilde t$ and $\tilde \chi_i^{\pm}, \tilde \chi_i^{0}$ are light. Light 
stop/charginos can decrease $\Gamma (\sh \to \gamma \gamma)$ 
through their contribution in the loop. For the light $\tilde t$ the inclusive 
production mode $gg \to \sh$ is also reduced substantially. If the channel 
$\sh \to \chi^0_1 \chi^0_1$ is open, that depresses the $BR$ into the 
$\gamma \gamma$ channel even further\cite{abdel1,abdel2,bbs,bbdgr}.
\begin{figure}[htb]
        \centerline{
             \includegraphics*[scale=0.50]{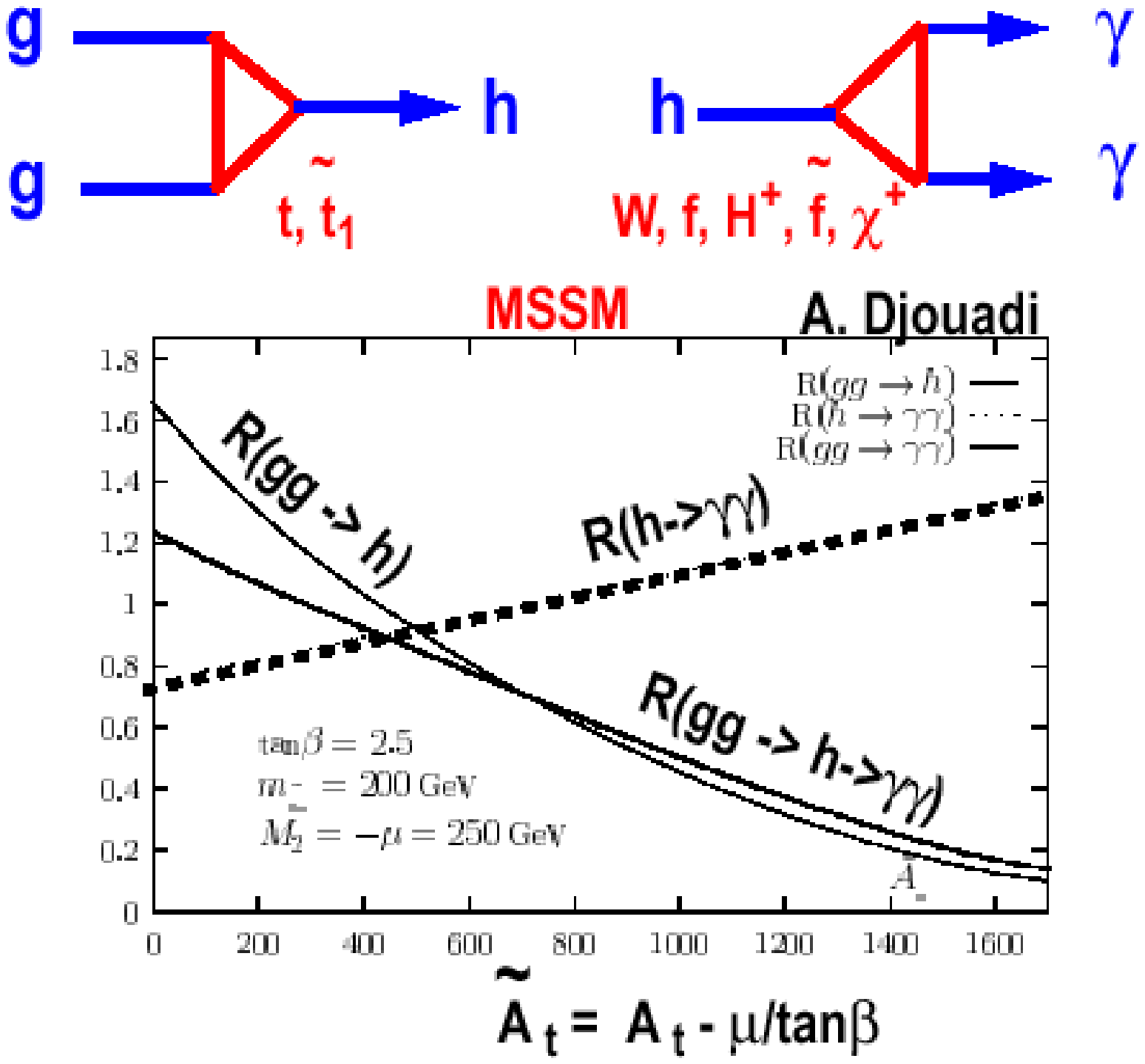}
              \includegraphics*[scale=0.30]{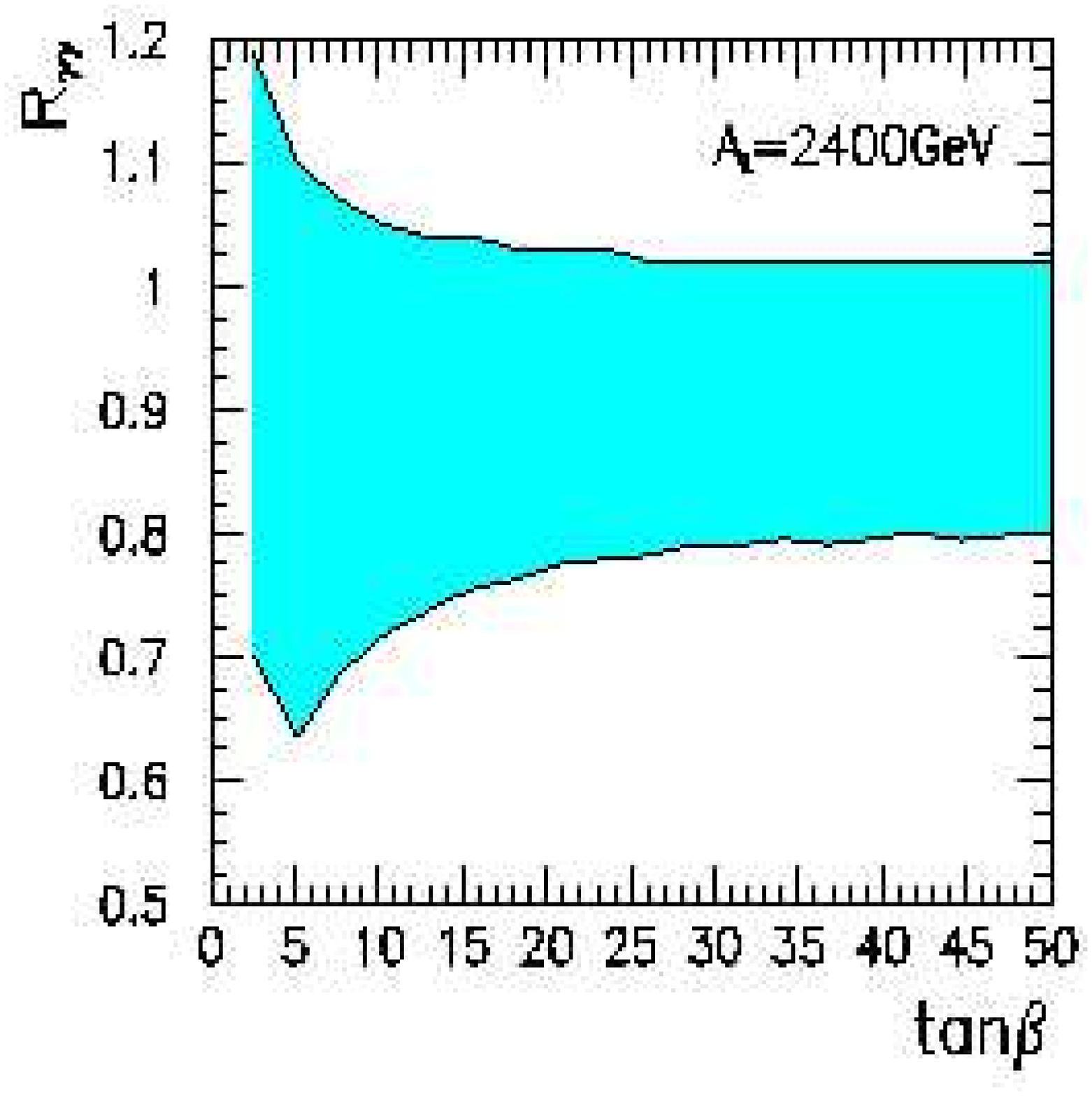}
}
\caption{Effect of light sparticles on the $\gamma \gamma$ decay
width  and $gg$ production of the Higgs\protect\cite{abdel1,bbdgr}.
\label{ffive}}
\end{figure}
The left panel in Fig.~\ref{ffive}\cite{abdel1} shows the ratio
$$R(h \to \gamma \gamma) = \frac{\Gamma (\sh \to \gamma \gamma)}
{\Gamma (h \to \gamma \gamma)}$$
and ratios $R(gg \to h), R(gg \to h \gamma \gamma)$ defined similarly.  Thus
we see that for low $\tan \beta$ the signal for the light neutral higgs
\sh\ can be completely lost for a light stop.  
The panel on the right in fig.~\ref{ffive}\cite{bbdgr}
shows $R(\sh \to \gamma \gamma)$ as a function of $\tan \beta$ for the case of
only a light chargino and neutralino. Luckily, eventhough light sparticles, 
particularly a light $\tilde t$ can cause disappearance of this signal, 
associated production of the  higgs \sh\ in the channel ${\tilde t}_1 
{\tilde t}^*_1 \sh /t \bar t \sh$ provides a viable signal. However, 
an analysis of the optimisation of the observability of such a light 
stop $(m_{\tilde t} \simeq 100-200 $ GeV) at the LHC still needs to be done.

\subsection*{Search for the MSSM higgs at \eplem\ colliders}
At an \eplem\ collider with $\rts \le 500$ GeV, more than one
of the MSSM Higgs scalar will be visible over most of the parameter
space~\cite{tesla,nlc,jlc,spira,abdel3}. 
At large $m_A$ (which seem to be the values preferred
by the current data on $b \to s \gamma$), the SM Higgs $h$  and \sh\ are
indistinguishable as far as their couplings are concerned.
Hence the most interesting question to ask is how well can one 
distinguish between the two. 
Recall from Fig.~\ref{ffour} that at the LHC there exists a largish region 
in the $\tan\beta - \ma$ plane where only the lightest \sh\ is observable if
only the SM--like decays are accessible.  With TESLA, the \sh\ boson can be 
distinguished from the SM Higgs boson through the accurate determination 
of its couplings and thus reveal its supersymmetric nature.
This becomes clearer in the 
\begin{figure}[hbt]
    \centerline{
 \includegraphics*[scale=0.40]{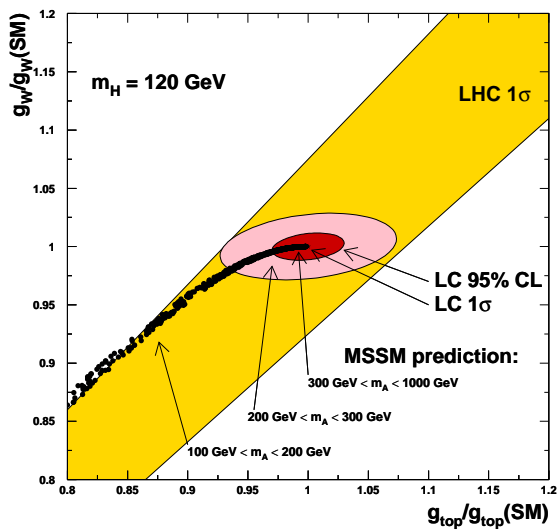}
}
\caption{ 
A comparison of the accuracy in the determination of the
$g_{tt\sh}$ and $g_{WW\sh}$ Higgs couplings at the LHC and 
at TESLA compared to the predictions from MSSM for different values 
of the \ma mass~\protect\cite{tesla}.}
\label{nlclhch2}
\end{figure}
Fig.~\ref{nlclhch2} which shows a comparison of the accuracy of the 
determination of the coupling of the $h$(\sh) with a $WW$ and $t \bar t$ pair,
for the LHC and TESLA along with expected values for the MSSM as a function
of $m_A$. 
It is very clear that the precise and absolute measurement of all 
relevant Higgs boson couplings can only be performed at TESLA.
The simplest way to determine 
the ${\cal CP}$ character of the scalar will be to produce $h$ in a \gamgam\ 
collider, the ideas for which are under discussion~\cite{tesla,nlc,jlc}.
An unambiguous determination of the quantum numbers of
the Higgs boson and the high sensitivity to ${\cal CP}$--violation
possible at such machines represent a crucial test of our ideas.
The measurement of the Higgs self coupling
gives access to the shape of the Higgs potential.
These measurements together will allow to establish the Higgs meachnism as the
mechanism of electroweak symmetry breaking. To achieve this goal in its 
entirety a linear collider will be needed~\cite{tesla,nlc,jlc}.

\section{Prospects for  SUSY search at colliders}
The new developements in the past years in the subject have been in trying
to set up strategies so as to disentangle signals due to different 
sparticles from each other and extract information about the SUSY breaking
scale and mechanism, from the experimentally determined properties and the 
spectrum of the sparticles~\cite{extrame}.
As we know, the couplings of  {\bf almost} all
the sparticles are determined by the symmetry, except for the charginos,
neutralinos and the light $\tilde t$. However, masses of all
the sparticles are completely model dependent, as has been already
discussed earlier. 
For
$\Delta M$ = $m_{\tilde{\chi}_{1}^{\pm}}$ - $m_{\tilde{\chi}_{1}^{0}}$
$<$ 1 GeV, the phenomenology of the sparticle searches in AMSB models will be
strikingly different from that in mSUGRA, MSSM etc. In the GMSB models, 
the LSP is gravitino and is indeed `light' for the range of the values of
$\sqrt{F}$ shown in Table~\ref{T:rgplen:1}. The candidate for the next lightest
sparticle, the NLSP can be $\N0_{1}$, $\stau_{1}$ or
$\sel_R$ depending on model parameters. The NLSP life times
and hence the decay length of the NLSP in lab  is given by 
$ L = c\tau \beta \gamma \propto \frac{1}{(M_{LSP})^5}$ $(\sqrt{F})^{4}$. 
Since the theoretically allowed 
values of $\sqrt{F}$ span a very wide range as shown in Table 1, so do those
for the expected life time and this range is given by 
$10^{-4}$ $<$ c$\tau \gamma \beta$ $<$ $10^{5}$ cm.
In first case the missing transverse energy 
\et\ is the main signal.  In the last case, along with \et\ 
the final states also have photons and/or displaced vertices, stable 
charged particle tracks etc. as the telltale signals of 
SUSY~\cite{tev_rep,atlas_tdr}.
In view of the lower bounds on the masses of the sparticle  masses established 
by the negative results at LEP, the most promising signal for SUSY at the run
II of the Tevatron, is the pair production of a chargino-neutralino pair 
followed by its leptonic decay giving rise to `hadronically quiet' trileptons.
The reach of Tevatron run-II for this channel, in mSUGRA is shown in 
\begin{figure}[htb]
\begin{center}
\includegraphics*[scale=0.4]{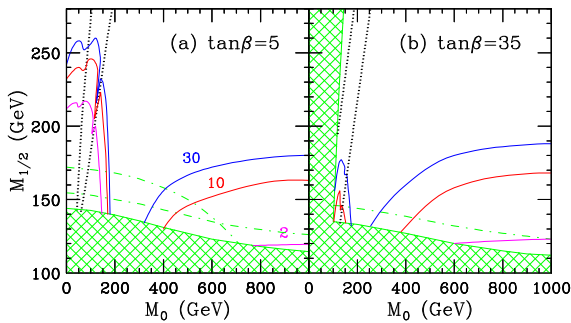}
\caption{\em Expected reach of the trilepton signal at the Tevatron 
run-II~ \protect\cite{matchev-3l}.}
\label{fsixp}
\end{center}
\end{figure}
Fig.~\ref{fsixp} taken from Ref.~\cite{matchev-3l}. It  shows the reach 
in the plane of mSUGRA parameters $M_0, M_{1/2}$ for two different 
values of $\tan\beta$. The dash-dotted lines correspond to the limits 
that have been reached by the latest LEP data. The left(right) dotted 
lines represent where the chargino mass equals that of the 
$\tilde\nu_{\tau} (\tilde \tau_R)$ for $\tan\beta =5$ and 
to $\tilde\tau_1 (\tilde e_R)$ for $\tan\beta = 35$.
\begin{figure}[htb]
\begin{center}
\includegraphics*[scale=0.4]{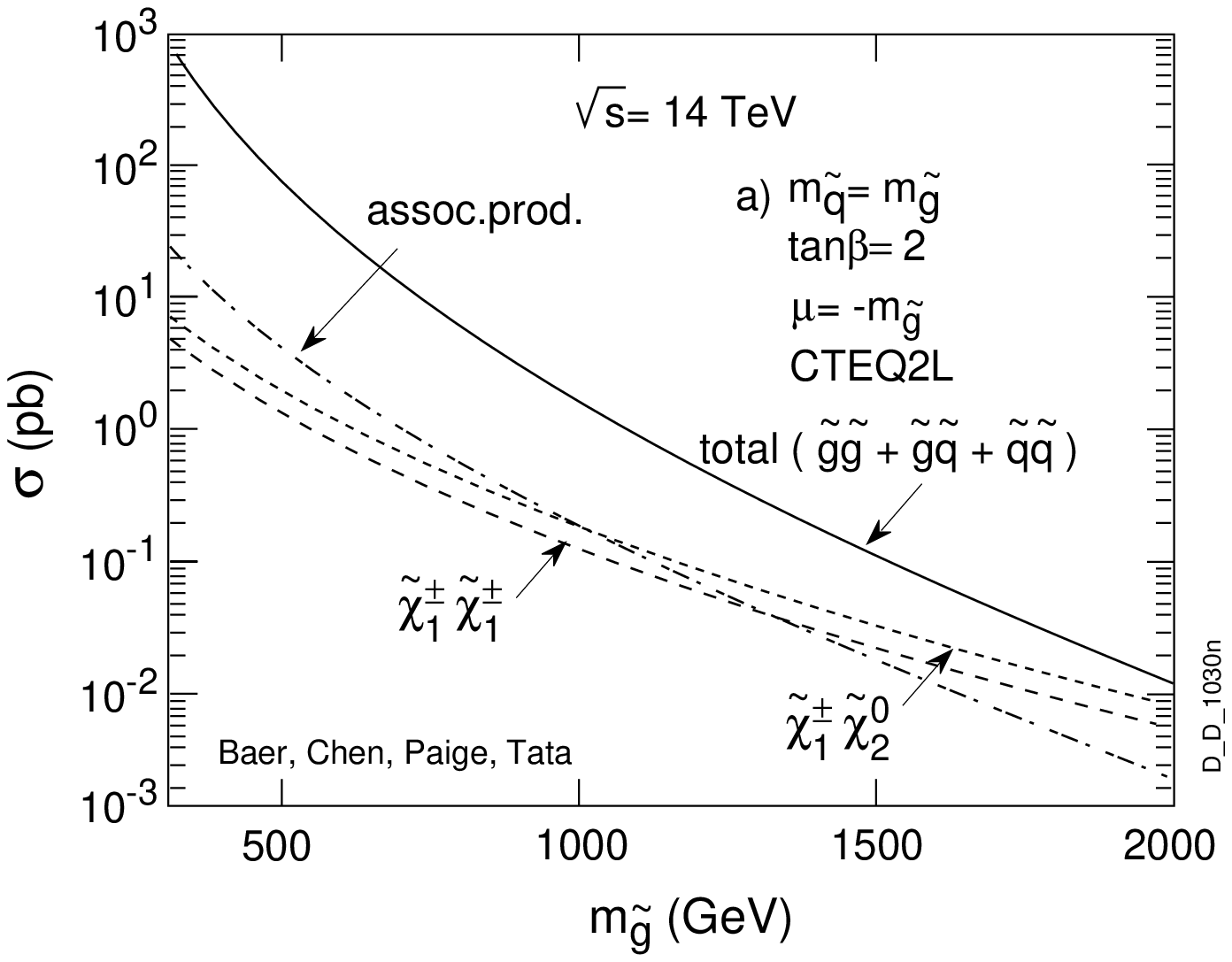}
\caption{\em Expected production cross-sections for various sparticles at
the LHC\protect\cite{tata}.}
\label{fsix}
\end{center}
\end{figure}

As is clear from the Fig.~\ref{fsix}, LHC is best suited for the search
of the strongly interacting $\tilde g, \tilde q$ because they have the 
strongest production rates.  The  ${\tilde\chi}_i^{\pm}, {\tilde \chi}^0_i$,
are produced via the EW processes or the decays of the $\tilde g, \tilde q$.
The former mode of production gives very clear signal of  `hadronically
quiet' events. The sleptons which can be produced mainly via the DY
process have the smallest cross-section. As mentioned earlier, various
sparticles can give rise to similar final states, depending on the mass
hierarchy. Thus, at LHC the most complicated background to SUSY search is
SUSY itself! The signals consist of events with \et, $m$ leptons and
$n$ jets with $m,n \geq 0$. Most of the detailed simulations which address the 
issue of the reach of LHC for SUSY scale, have been done in the context of 
mSUGRA picture. 
\begin{figure}[htb]
\begin{center}
\includegraphics*[scale=0.4]{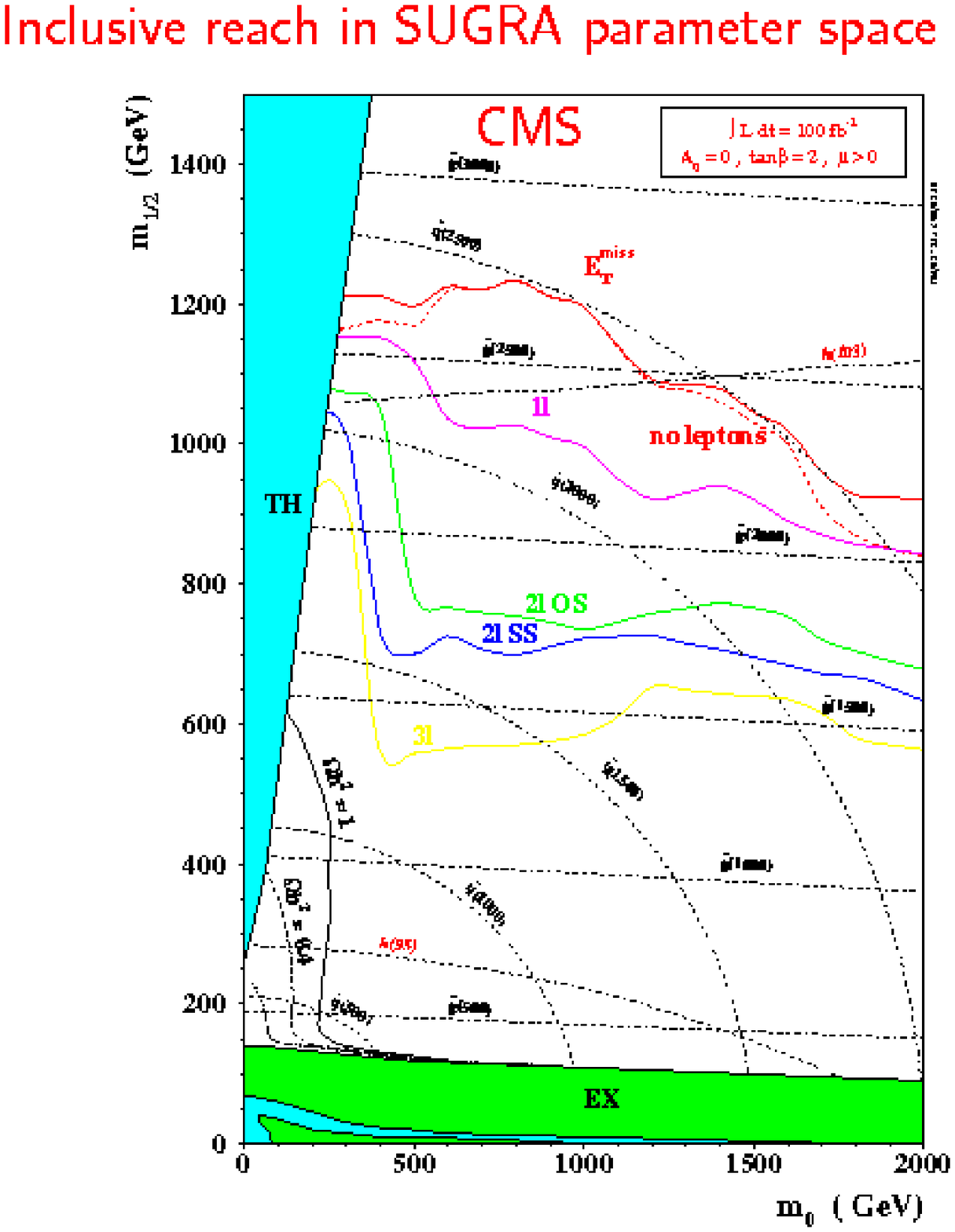}
\caption{\em Expected reach for SUSY searches at the LHC\protect\cite{polsel}.\label{fseven}}
\end{center}
\end{figure}
We see from fig.\ref{fseven} that for $\tilde g, \tilde q$ the
reach at LHC is about $2.5$ TeV and over most of the parameter space 
multiple signals are observable. Note that $m_0,m_{1/2}$ used in this figure 
are the same as $M_0,M_{1/2}$ used in the text and other figures.

To determine the SUSY breaking scale $M_{SUSY}$ from the jet events, a method 
suggested by Hinchliffe et al\cite{paige} is used, which consists in defining
$$M_{eff}  =  \sum_{i=1}^4 |P_{T(i)}| + \et$$ 
and looking at the distribution in $M_{eff}$. The jets, that are produced by 
sparticle production and decay, will have
$P_T \propto m_1 - \frac{m_2^2}{m_1}$, where $m_1,m_2$ are the masses of the
decaying sparticles. Thus this distribution can be used to determine 
$M_{SUSY}$.
\begin{figure}[htb]
\begin{center}
\includegraphics*[scale=0.4]{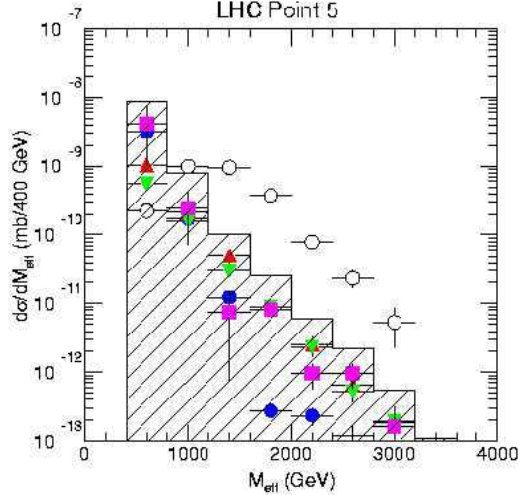}
\caption{\em  Determination of SUSY breaking scale using jet events at 
LHC point \protect\cite{paige}.\label{feight}}
\end{center}
\end{figure}
The distribution in fig.~\ref{feight} shows that indeed there is 
a shoulder above the SM background. The scale $M_{SUSY}$ is defined either 
from the peak position or the point where the signal is aprroximately equal to the
background. Then of course one checks how well $M_{SUSY}$ so determined
tracks the input scale. A high degree of correlation was observed in
the analysis, implying that this can be a way to determine the SUSY breaking 
scale in a precise manner. 

It is possible to reconstruct the masses of the charginos/neutralinos
using kinematic distributions.
\begin{figure}[htb]
\begin{center}
\includegraphics*[scale=0.4]{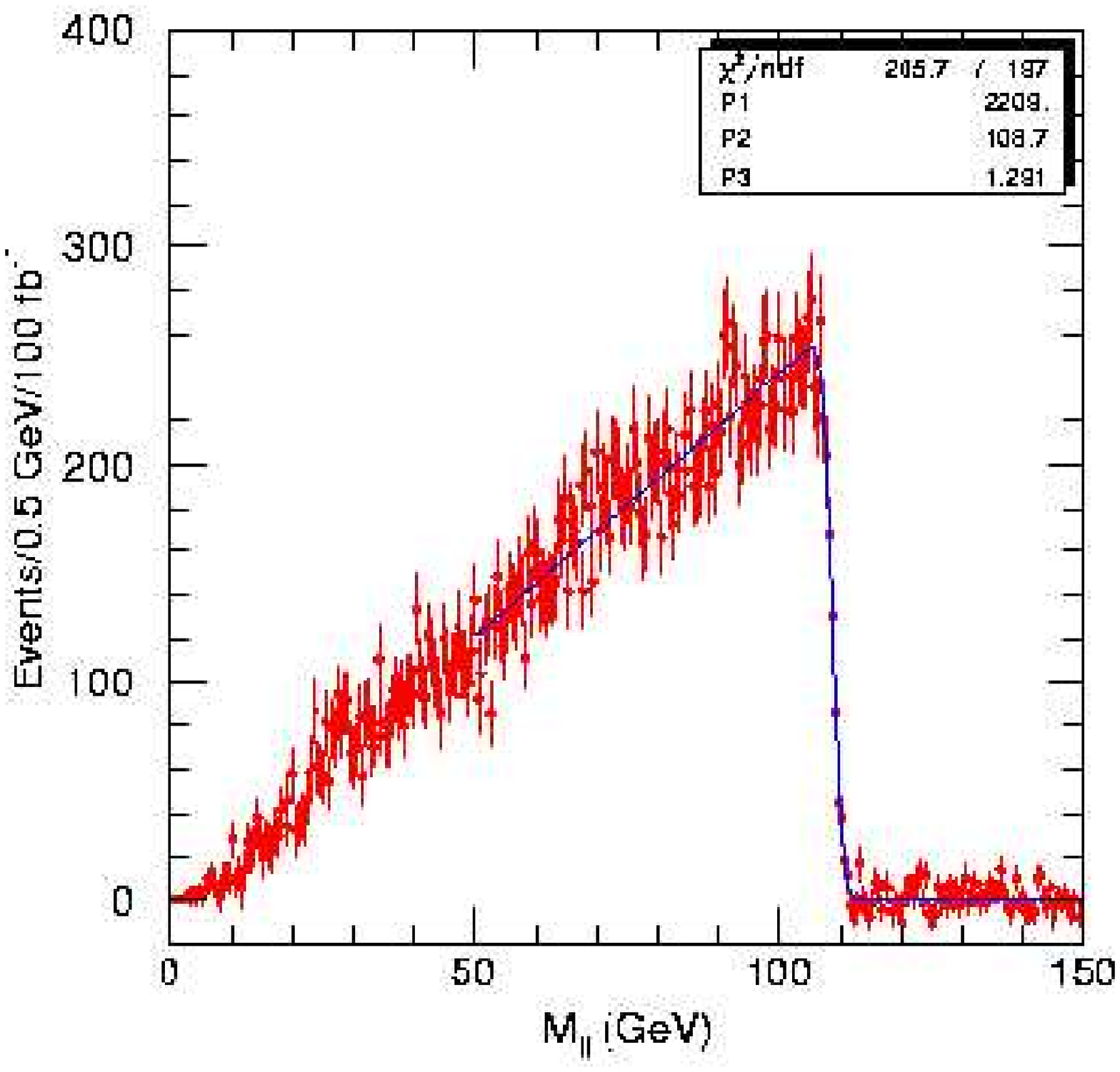}
\caption{\em Kinematic reconstruction of the mass of ${\tilde\chi}_2^0$ 
from the dilepton mass distribution~\protect\cite{polsel}.}
\label{fnine}
\end{center}
\end{figure}
Fig.~\ref{fnine} demonstrates this, using the distribution in the
invariant mass $m_{l^+l^-}$ for the $l^+l^-$ pair produced in the
decay ${\tilde \chi}^0_2 \to {\tilde \chi}_1^0 l^+ l^-$. The end 
point of this distribution is $\sim m_{{\tilde\chi}_2^0} - m_{{\tilde \chi}_1^0}$. However, such analyses have to be performed with caution. As 
pointed out by
Nojiri et al\cite{mihoko1}, the shape of the spectrum near the end point can
at times depend very strongly on the dynamics such as the composition of the
neutralino and the slepton mass. One can still use these determinations
to extract model parameters, but one has to be careful.

\subsection*{Studies of SUSY and SUSY breaking scale at the \eplem\ colliders}
At the \eplem\ colliders one can study with great accuracy the production 
of sleptons, squarks, charginos and neutralinos. The precision measurements 
of the sparticle masses and quantum numbers allows testing the basic 
predictions of supersymmetry about the particle spectrum.
There have been a large number of dedicated studies of the possibilities 
of precision measurements of the properties of these sparticles 
at the next generation of Linear Colliders~\cite{MP1,extra3,refall}. 
Study of the third generation of sfermions are shown 
to yield particularly interesting information about SUSY models. Almost 
no study is possible without use of polarisation, at least for one of the  
initial state fermions.

\noindent\underline{{\it Sleptons and Charginos/Neutralinos}}

\noindent The masses of sleptons can be determined at an LC essentially using
kinematics. Making use of partial information from the LHC, 
it will be
possible to tune the energy of the LC to produce the sfermions sequentially.
The pair produced lightest sleptons will decay through a two body decay.
Let us take the example of $\smu_R$ which will have the simplest decay.
So one has in this case, 
$$
e^+e^- \rightarrow \smu_R \smu_R^{*} \rightarrow \mu^+ \mu^-
\tilde{\chi}_{1}^{0} \tilde{\chi}_{1}^{0}
$$
Since the slepton is a scalar the decay  energy distribution for the
$\mu$ produced in two body decay of $\smu_R$, will be flat with 
$$
{m_{\smu_R} \over 2} \left(1-{{m_{\N0_1}^2} \over {m_{\smu_R}^2}} \right)
 \gamma (1-\beta) < E_\mu <
{m_{\smu_R} \over 2} \left(1-{{m_{\N0_1}^2} \over {m_{\smu_R}^2}}\right)
\gamma (1+\beta) .
$$
Thus measuring the end points of the $E_{\mu}$ spectrum accurately will
yield a precision measurement of the masses $m_{\smu_{R}}$,
$m_{\N0_1}$~\cite{tesla,nlc,jlc}.  However, the method of using the end 
point of the energy spectrum will not work so well for the third 
generation slepton,\eg ,  for $\stau_{1} \stau_{1}^{*}$ production and decay
~\cite{MIHO2,tata3}.

Another method for precision determination of the masses of the sleptons and 
the lighter charginos/neutralinos, is to perform threshold scan. The 
linear $\beta$
dependence as opposed to the $\beta^{3}$ dependence of the cross-section,
near the threshold (where $\beta$ is the c.m. velocity of the produced
sparticle) makes the method more effective for the spin 1/2
charginos/neutralinos than the sleptons. The threshold scans offer the
possibility of very accurate mass determinations~\cite{tesla} of the
sparticles, albeit with very high luminosity.

Recent analyses of the mass determination of $\stau_{1}$~ and
${\snu}_{\tau_{1}}$~\cite{tata3}  using the continuum production, 
show that only an accuracy of $\sim 2\%$ 
for $m_{\stau_1}$ (consistent with the earlier analyses~\cite{MIHO2}) 
and even much worse $6-10 \%$ for $m_{\snu_\tau}$, is possible even after 
a use of optimal polarisation and comparable luninosities as in the 
threshold scan  case.  A recent study~\cite{tata4} addressing  these issues 
indicates that
threshold scans may not be the optimal way to measure the masses 
for the second and third generation sleptons. This is caused by the low rates
which force one to go away from the threshold and also the `a priori'
unknown branching ratios of the $\stau_1$ or $\snu_\tau$.
It is very important to clearly understand
just how well these measurements can be made, as these accuracies 
affect,  crucially, the projected abilities to glean information about the SUSY 
breaking scale.
 

\vspace{0.5cm}

\noindent\underline{{\it Squarks}}

\vspace{0.2cm}

\noindent Clearly squarks are the only strongly interacting particles about
whom direct information can be obtained at the  $e^{+}e^{-}$ collider. For
the strongly interacting sfermions (squarks) the decay is $\sq \to
q \N0_{1}$. As a result, one has to study the end point of the
distribution in $E_{jet}$. The hadronization effects can in principle 
deteriorate
the accuracy of the determination of $m_{\sq}$. An alternate estimator
~\cite{FF} of  $m_{\tilde{q}}$ is the peak of the distribution in the minimum
kinematically allowed mass of the $q\N0_{1}$ system produced in
\begin{figure}[ht]
\centerline{
 \includegraphics*[scale=0.4]{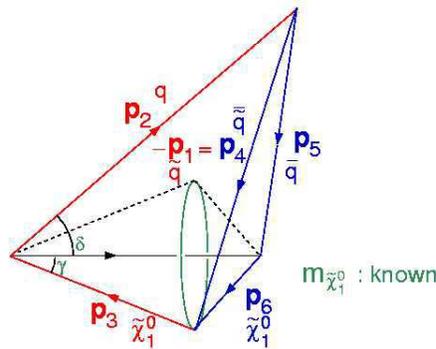}
}
\caption{Determination of the minimum kinematically allowed squark mass, 
following Ref.~\protect\cite{FF}.}
\label{F:rgplen:4}
\end{figure} 
$\tilde{q}$ decay; $m_{\tilde{q}, min}$. The minimum squark mass corresponds 
to maximum possible $|\vec p_4|$ and can be easily determined following the
construction in Fig.~\ref{F:rgplen:4}. The figure in the left panel of
Fig.~\ref{F:rgplen:5}, taken from Ref.~\cite{FF} shows the efficacy of 
this estimator for a 500 GeV machine with $10$ fb$^{-1}$ luminosity per 
polarisation, the latter 
\begin{figure}[htb]
\centerline{
 \includegraphics*[scale=0.40]{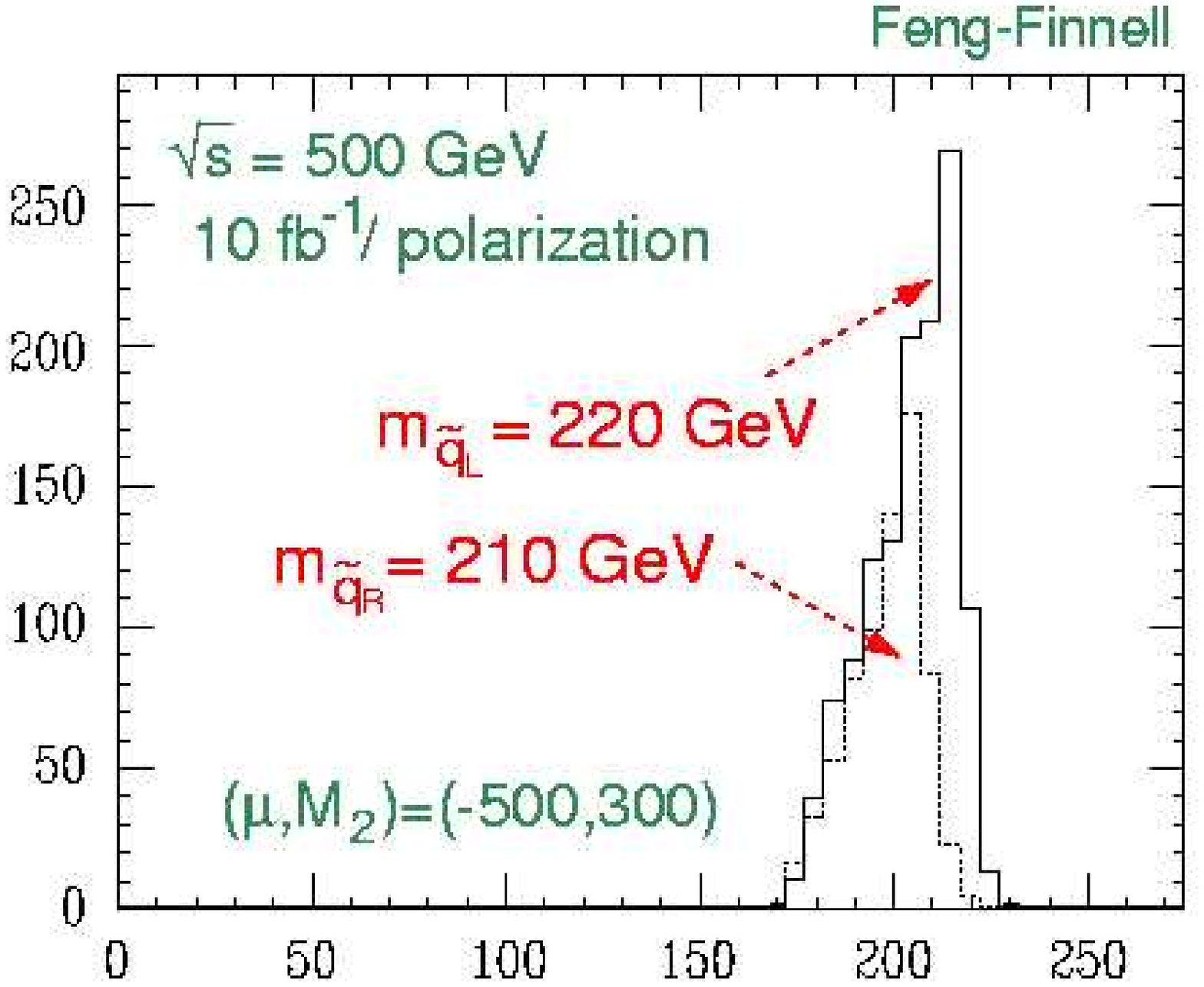}
 \includegraphics*[scale=0.40]{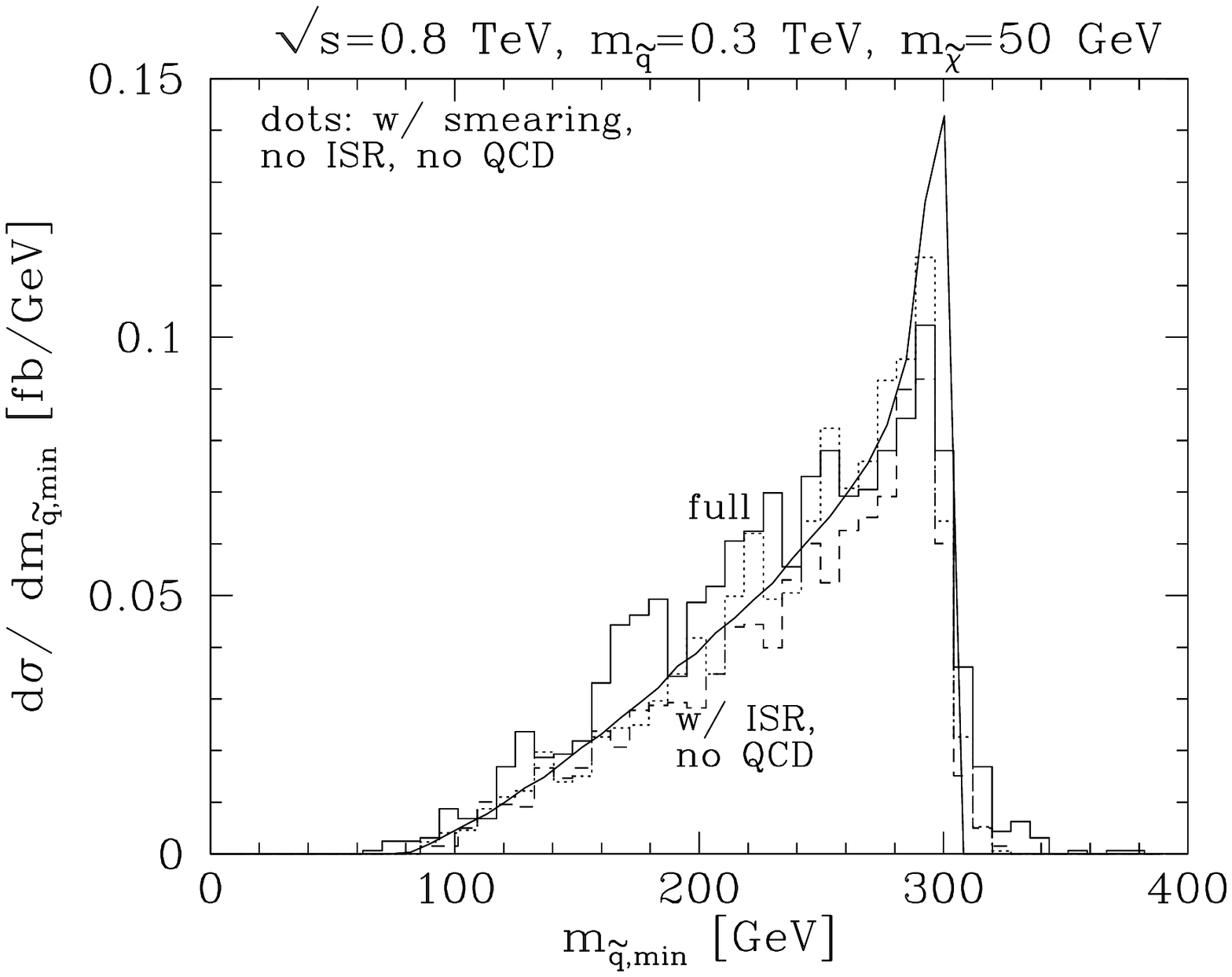}
}
\caption{Accuracy of determination of $m_{\sq}$ using the $m_{\sq,min}$ defined 
in~\protect\cite{FF}.}
\label{F:rgplen:5}
\end{figure} 
being used for separating $\sq_{L}/ \sq_{R}$ contributions, for
a particular point in the MSSM parameter space. Figure in the right panel
shows that this variable provides a good estimate of  $m_{\sq}$ even
after radiative corrections, both in production and decay, have been
included~\cite{DEKG}.

If the squarks are lighter than the glunios, any information on gluino 
masses at an  $e^{+}e^{-}$ collider can only come from the assumed relations 
between the masses of the electroweak and strong gauginos.

\subsection*{Precision determination of mixings}
The mixing between various interaction eigenstates in the gaugino sector as
well as the, in general, large mixing in the L-R sector for the third
generation squarks and sleptons, is decided respectively by $M_1, M_2, \mu,
\tan\beta$ and $\mu, \tan\beta, A $ as well as various scalar mass
parameters. So clearly an accurate measurement of these mixings along with
the precision measurements of masses offers further clues to physics at
high scale. 

\begin{figure}[htb]
\centerline{
 \includegraphics*[scale=0.80]{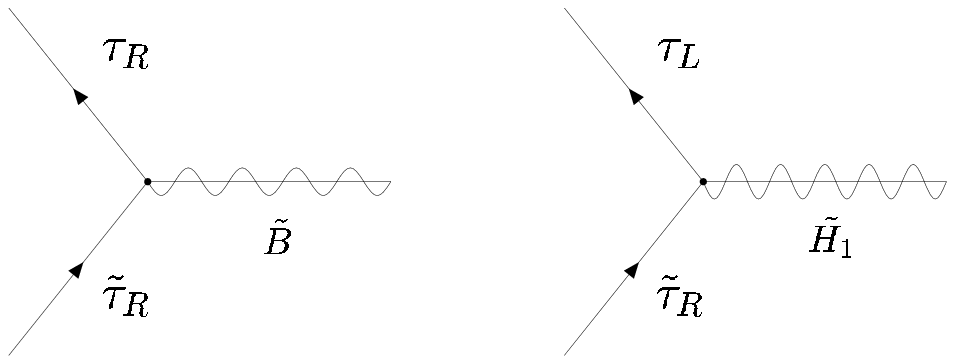}
}
\caption{ $\stau_{R} \to \tau \N0_{1}$. }
\label{F:rgplen:6}
\end{figure}

Possibilities of the determination of L-R mixing in the third generation
sfermions have been investigated~\cite{tesla,nlc,jlc}. The mass eigenstates 
can be written
down in terms of the interaction eigenstates for,\eg, staus as
$\stau_{1} = \stau_{L} \cos \theta_{\tau} + \stau_{R}
\sin \theta_{\tau}$,$\stau_{2} = \stau_{L} \sin \theta_{\tau}
+ \stau_{R} \cos\theta_{\tau}$.
It is clear that polarised $e^{-}$/$e^{+}$ beams can play a crucial role in
determining  $\theta_{\tau}$. Let us, for example, consider
$e^{+} e^{-} \to \stau_{1} \stau_{1}^{*}$. Further
let us consider the case of 100 \% polarisation in particular. The pair
production proceeds through an exchange of $\gamma$/Z in $s$-channel. For
energies $\sqrt{s}$ $>$ $>$ $m_{Z}$, with $P_{e^{-}}$ = 1, one can
essentially interpret this $s$-channel exchange of  $\gamma$/Z as an U(1)
gauge boson, $B$.  In this limit $\sigma (\stau_{R})$ = 4
$\sigma (\stau_{L})$. Thus it is clear that a measurement of $\sigma
( e^{+} e^{-} \to  \stau_{1}^{*}  \stau_{1} )$ along with a
knowledge of polarisation of $e^{-}$ beam can lead to an extraction of $\cos
\theta_{\tau}$. Further the polarisation of $\tau$ produced in
$\stau_{1}$ decay provides a measurement of the mixing angle in the
neutralino sector as well~\cite{MIHO1}. Let us consider $\stau_{R} \to \tau
\N0_{1}$ depicted in Fig.~\ref{F:rgplen:6}. The $\tilde{B}$
component of $\N0_{1}$ produces  $\stau_{R}$, whereas
the higgsino
component will flip the chirality and produce $\stau_{L}$. Thus the
measurement of $ \stau_{1}^{*}  \stau_{1}$ production with
polarised $e^{-}$ beams and the polarisation of decay $\tau^{s}$ can 
give very useful information on both the mixings: the $L-R$ mixing in the
stau sector and the mixing in the neutralino sector. The $\tau$ polarisation 
can be measured by looking at the energy distribution of the decay 
product $\rho$ in the hadronic decay of $\tau$~\cite{MIHO1}.
\begin{figure}[ht]
\centerline{
 \includegraphics*[scale=0.35]{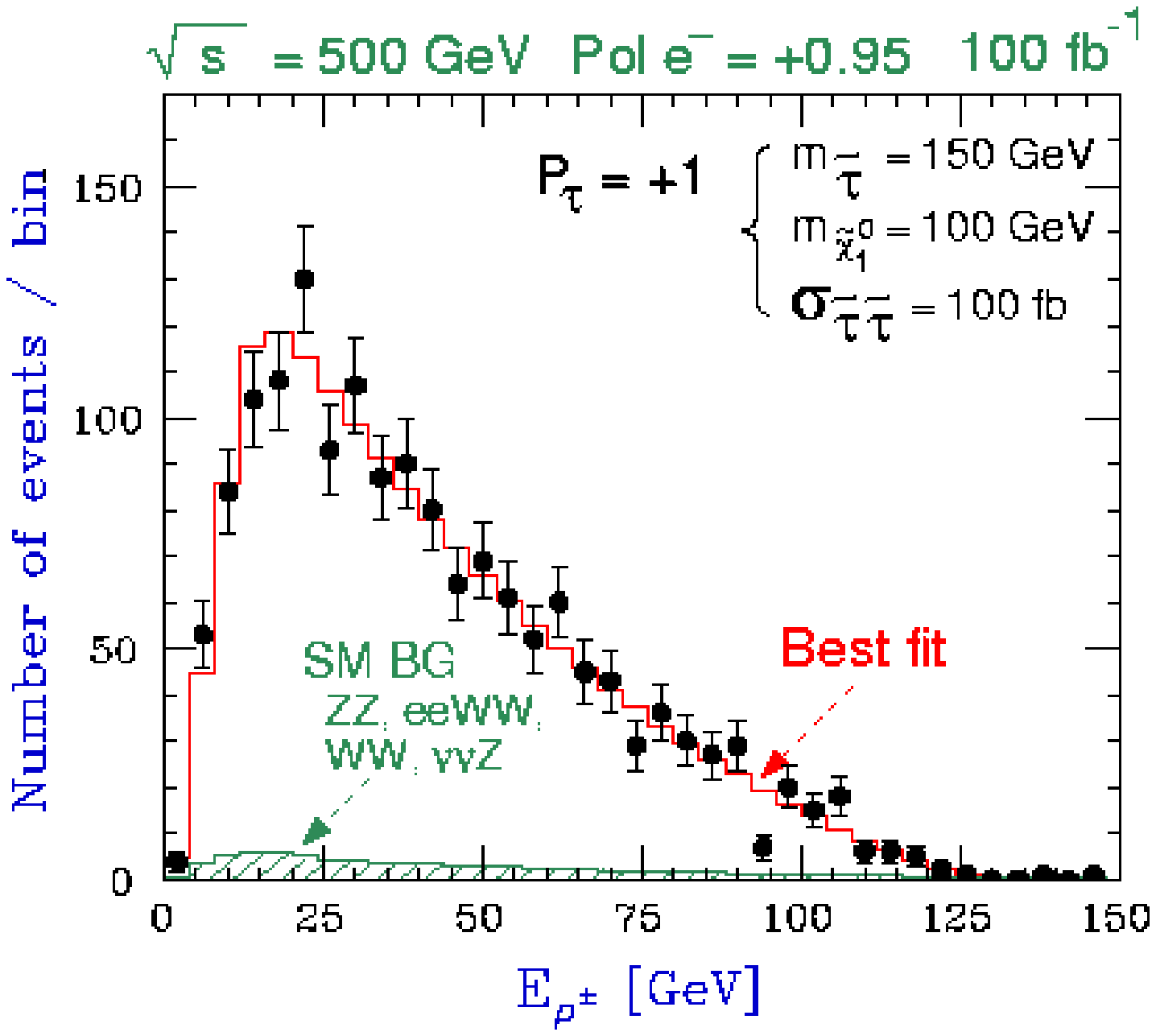}
 \includegraphics*[scale=0.35]{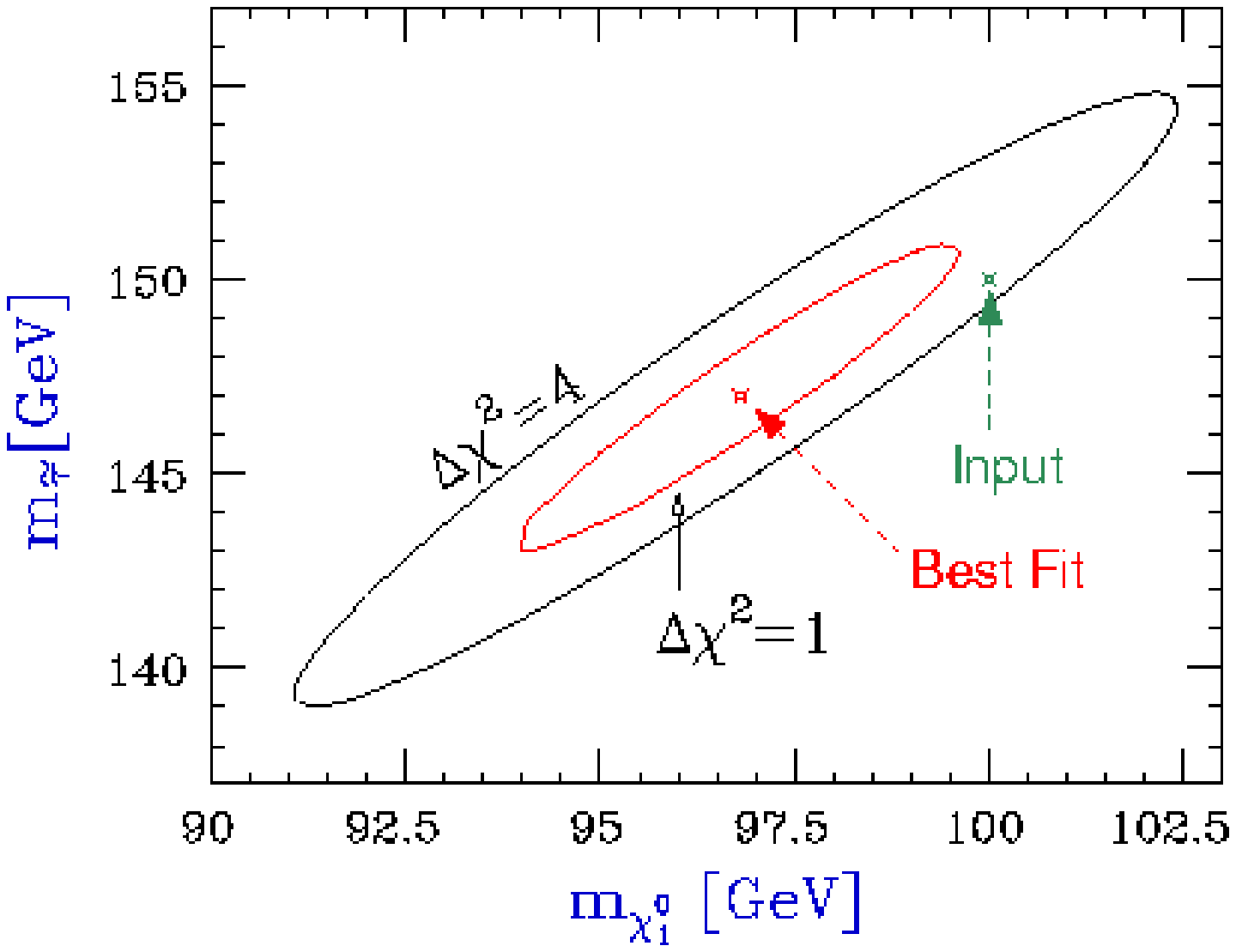}
}
\caption{Precision of the determination of $m_{\stau_{1}}$ and
$m_{\N0_{1}}$ for luminosities and polarisation 
mentioned in the figure. Taken from Ref.\protect\cite{MIHO2}.}
\label{F:rgplen:7}
\end{figure} 
Fig.~\ref{F:rgplen:7}~\cite{MIHO2} shows the possible accuracy of a
simultaneous determination of $m_{\tilde{\tau_{1}}}$ -
$m_{\tilde{\chi_{1}^{0}}}$ from the determination of the end points of the
energy spectrum, for $\int {\cal L} dt = 100\ {\rm fb}^{-1},P_{e^{-}} = 0.95$ 
and $\sqrt{s} = 500$ GeV. The input value lies outside the $\Delta
\chi^{2}$ = 1 contour around the best fit value. However, if
$m_{\N0_{1}}$ is assumed to be known, then  $\Delta m_{\stau_{1}}$ goes 
down considerably and a 1-2\% determination at 
$1 \sigma$ level is possible. 

\begin{figure}[htb]
\centerline{
 \includegraphics*[scale=0.35]{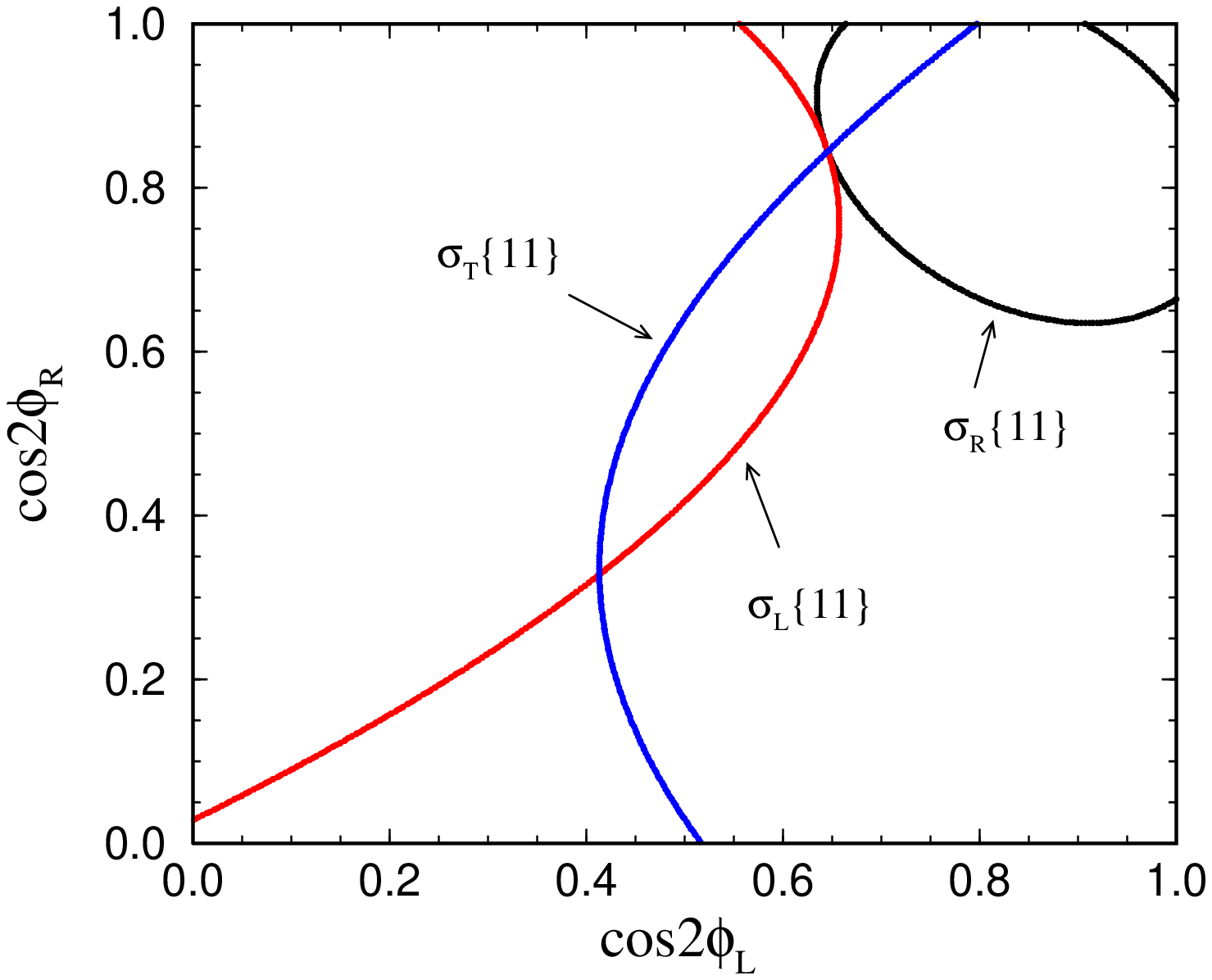}
}
\vspace {0.20in}
\centerline{
 \includegraphics*[scale=0.50]{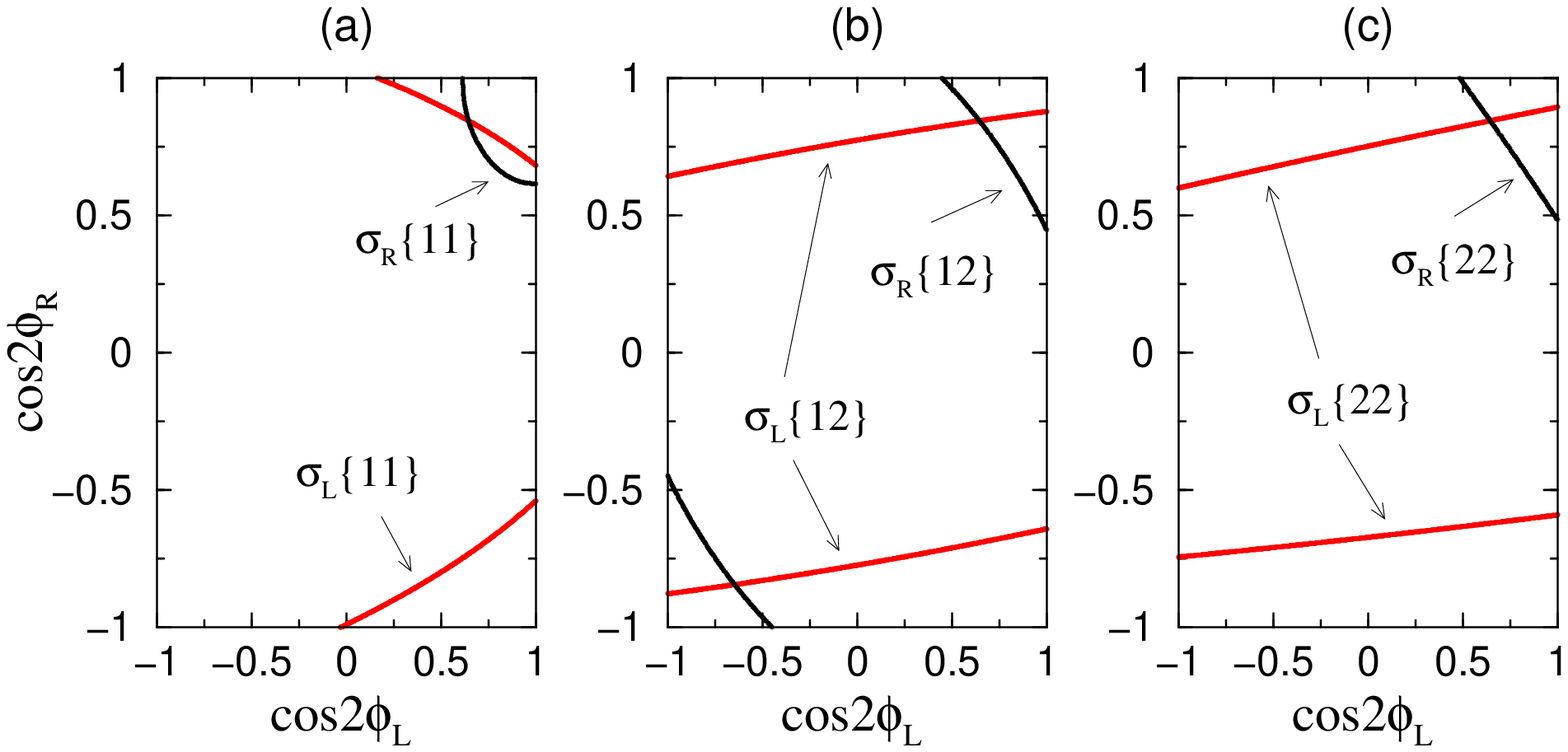}
}
\caption{Demonstration of unique determination of mixing angles in the
chargino sector using polarised beams~\protect\cite{Z2}.}
\label{F:rgplen:10}
\end{figure}

A study of the chargino sector at the LC can provide a precision determination
of the higgsino-gaugino mixing and consequently an accurate determination
of {\bf all} the Lagrangian parameters which dictate the properties 
of the chargino sector~\cite{jlc1,jlc2,Z2,tesla,nlc}. This requires, 
along with  the determination of 
$m_{\Chipm_1}, m_{\N0_1}$ and $\sigma^{\rm tot}_{\Chip_1,\Chim_1}$, 
a study of either the dependence of the production cross-section on 
the initial beam polarisations or the polarisation of the produced charginos 
through the angular distribution of their decay products. Since 
$\sigma(\epem \to \Chim_i \Chip_j)$ depends on  $m_{\snu}$,
its knowledge is also necessary. This 
can be obtained by studying the energy dependence of the $\sigma^{\rm tot}$,
even if $m_{\snu}$ is beyond the kinematic range of the collider. If only 
the lightest chargino is available kinematically, then one can determine the
mixing angles in the chargino sector $\Phi_L, \Phi_R$ defined through 
\begin{equation}
\Chim_{1L} = \cos \Phi_L \tilde W^{-}_{L} + \sin \Phi_L \tilde H^-_{2L},
\Chim_{1R} = \cos \Phi_R \tilde W^-_R + \sin \Phi_R \tilde H^-_{1R},
\end{equation}
only upto a two fold ambiguity.  However, this can be removed, 
using the information on the transverse polarisation, as shown in               
Fig.~\ref{F:rgplen:10}. If both the charginos are accessible kinematically,
$\cos 2\Phi_R, \cos 2\Phi_L$ can be determined uniquely through
measurements of $\sigma_{L/R} (\Chip_i \Chim_j)$, as shown in the lower panel 
of the same figure. It has been shown, in a purely theoretical study
~\cite{Z2}, in the context of TESLA, using only statistical 
errors, that with
$\int {\cal L} dt = 1\ ab^{-1}, \cos 2\Phi_L, \cos 2\Phi_R$ can be determined
to an accuracy of $\sim 1-3 \% $. Along with the information on $m_{\Chip_i}$
the mixing angles can then lead to an unambiguious determination of the
Lagrangian parameters $M_2,\mu$ and $\tan \beta$. However, since all the
variables are proportional only to $\cos 2\beta$, the accuracy of $\tan \beta$
determination is rather poor at high $\tan \beta$. At high $\tan \beta$,
measurements in the slepton sector (stau/selectron) discussed 
earlier~\cite{MIHO2} afford a better determination. 

\begin{figure}[htb]
\centerline{
 \includegraphics*[scale=0.60]{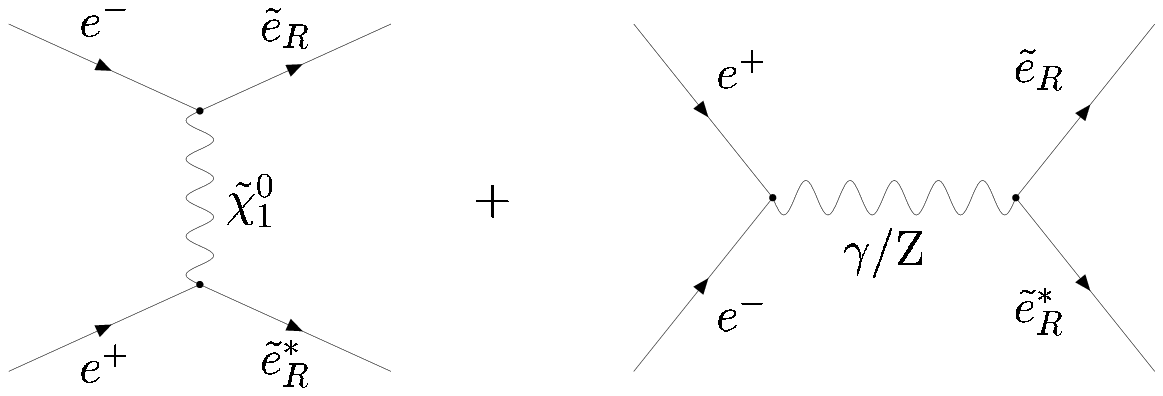}
 \includegraphics*[scale=0.30]{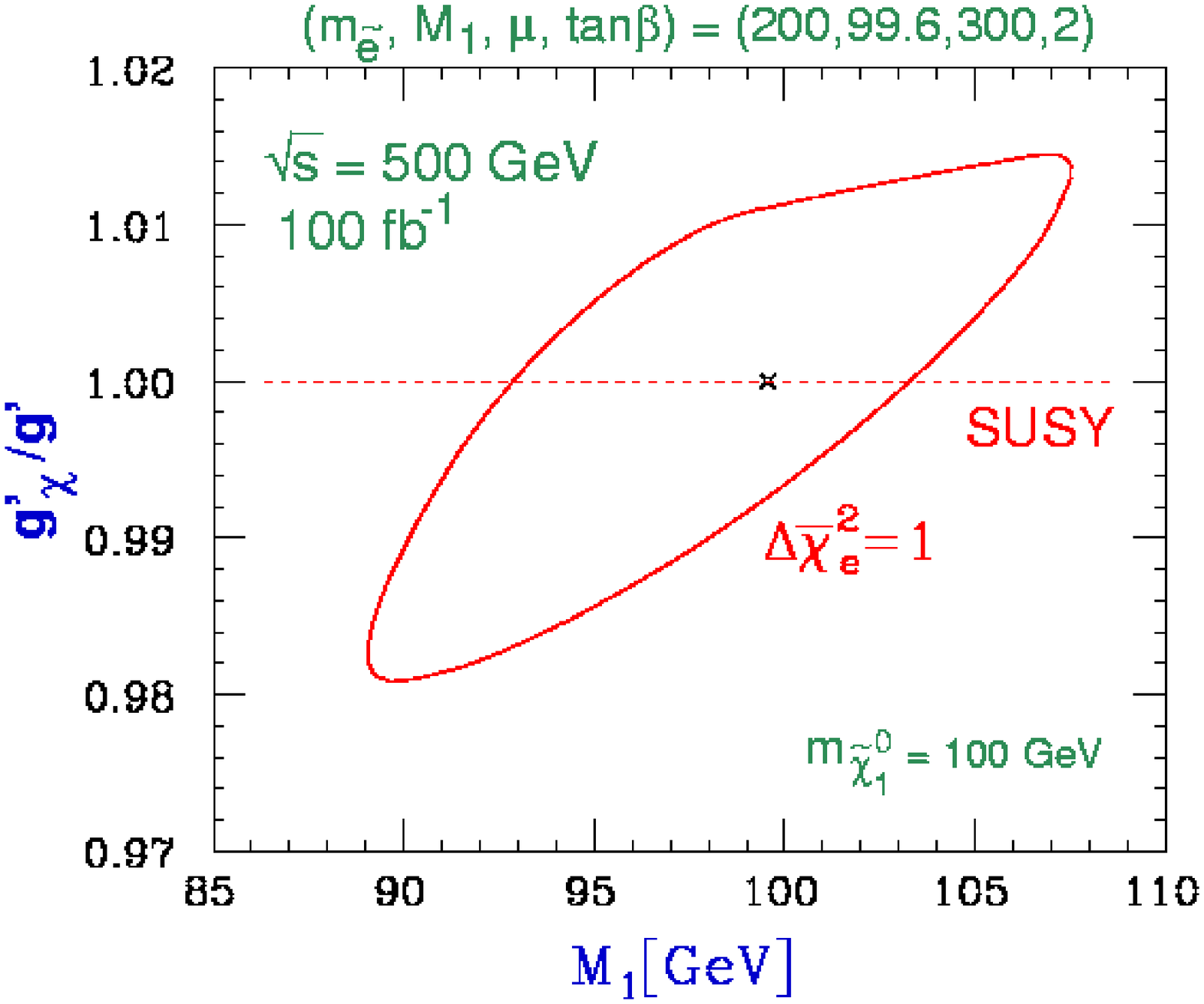}
}

\caption{Simultaneous determination of $M_1$ and $g_{\sel_R e \widetilde B}$,
in a study of $\sel_R \sel_R^*$  production.}
\label{F:rgplen:12}
\end{figure}

Above discussion already shows how an efficient use of polarisation of both
$e^+/ e^-$ beams, allows a high precision determination of the mixings 
among the $L-R$ sfermions as well as in the gaugino-higgsino sector. This is,
indeed, indirectly a determination of the hypercharges of the various 
sparticles. It has been demonstrated~\cite{jlc2}, using realistic 
simulation of the
backgrounds, that it is possible to reconstruct the $\smu$ angular distribution 
in the process $\epem \to   \smu \smu^* \to \mu^+ \mu^-  + \mET $ \  and hence 
determine the spin of the smuon with precision. Further, the cross-section
of $\sel_R \sel_R^* $ production can be used as a very sensitive probe of
the equality of the couplings $g_{\sel e_R \tilde B}$ and $g_{eeB}$. This is 
due the contribution of the $t$ channel diagram shown in the left-hand panel
of Fig.~\ref{F:rgplen:12}, which involves a $\N0_i$ exchange. The contribution 
to the production cross-section of the  $\sel_R \sel_R^*$ pair is
sensitive  to the bino component of $\N0_i$ and hence to the $U(1)$ gaugino 
mass parameter $M_1$ and the coupling $g_{\sel_R e_R B}$. At tree level 
we expect, due to supersymmetry, 
\begin{equation}
g_{\tilde{e}_R e_R \tilde B} = g_{eeB}  = \sqrt{2} g_2 \tan\theta_W =
\sqrt{2} g_1 Y_b = g_Y.
\label{eq:rg:1}
\end{equation}
Using $\sigma (\epem \to \sel_R \sel_R^* ) $ 
and $\frac{d\sigma}{d\cos\theta}
(\epem \to \sel_R \sel_R^*) $, 
one can determine $g_{\sel_R e_R \tilde B}$ and
$M_1$ simultaneously. For an integrated luminosity of $100\ {\rm fb}^{-1}$,
$Y_b$ of Eq.~\ref{eq:rg:1} can be determined to an accuracy of 
$1\%$~\cite{MIHO2}. This is shown in  the right panel of 
Fig.~\ref{F:rgplen:12}. This expected accuracy is actually 
comparable to the size of the SUSY radiative corrections~\cite{MIHO3}
 to the tree
level equality of Eq.~\ref{eq:rg:1} and hence this measurement can serve as
an indirect probe of the mass of the heavy sparticles. 

The precision measurements of the masses and the mixings in the sfermion and
the chargino/neutralino sector at the LC will certainly allow to establish
existence of supersymmetry as a dynamical symmetry of particle interactions.
However, this is not all these measurements can achieve. The high precision 
of these measurements will then allow us to infer about the SUSY breaking scale
and the values of the SUSY breaking parameters at this high scale, just the
same way the high precision measurements of the couplings $g_1,g_2$
and $g_3$ can be used to get a glimpse of the physics of unification and its
scale. 

There are essentially two different approaches to these studies. In the 
pioneering studies~~\cite{jlc2,MIHO2}, the JLC group investigated how 
accurately one can determine the parameters $M_1,M_2,\mu,\tan\beta$ and $M_0$ 
{\it at the high scale} by fitting these {\bf directly} to the various 
experimental observables such as the polarisation dependent production
cross-sections of the sparticles, angular distributions of the decay products 
etc., that have been mentioned in the discussion so far. 
\begin{figure}[htb]
\centerline{
 \includegraphics*[scale=0.35]{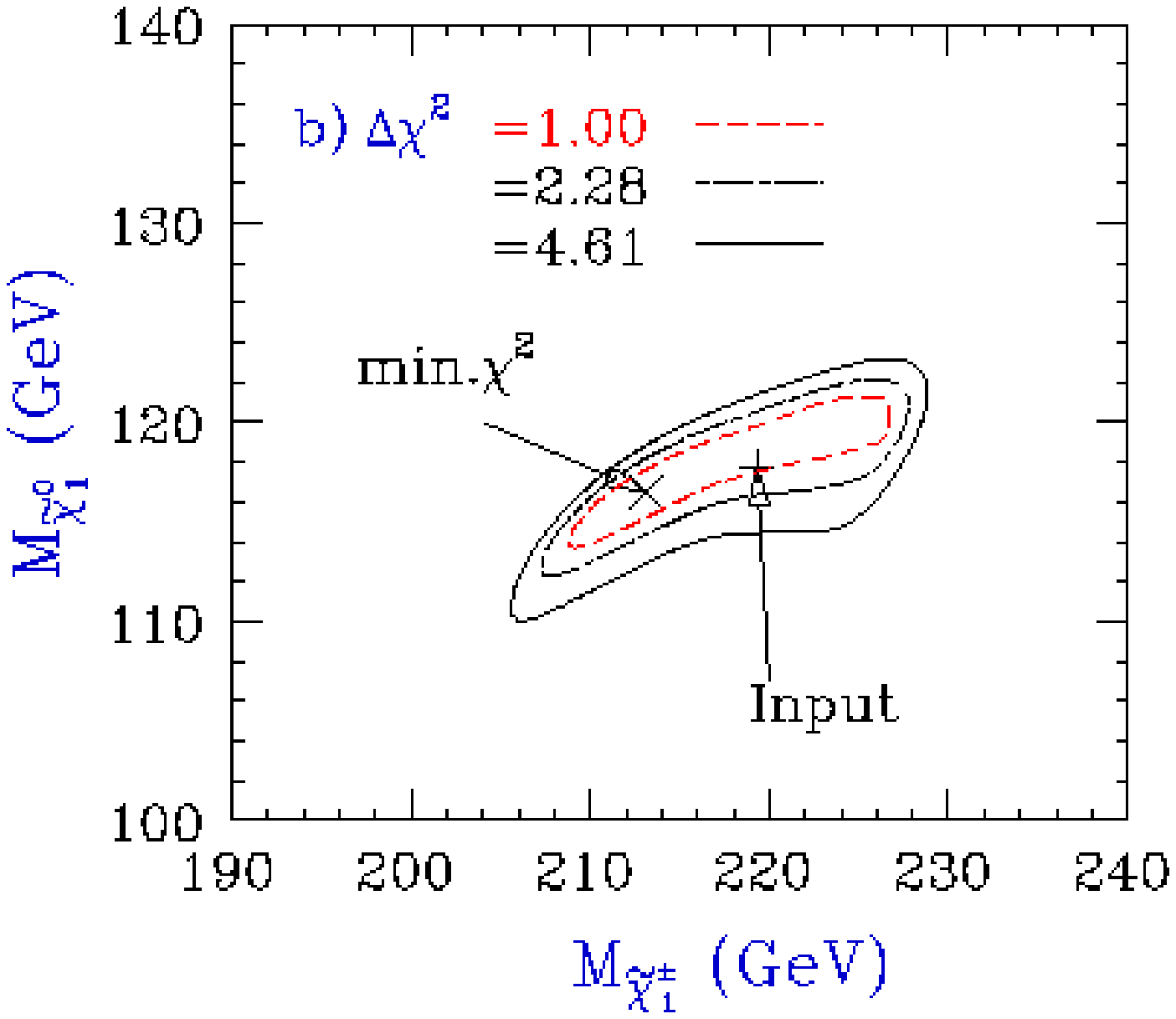}
 \includegraphics*[scale=0.35]{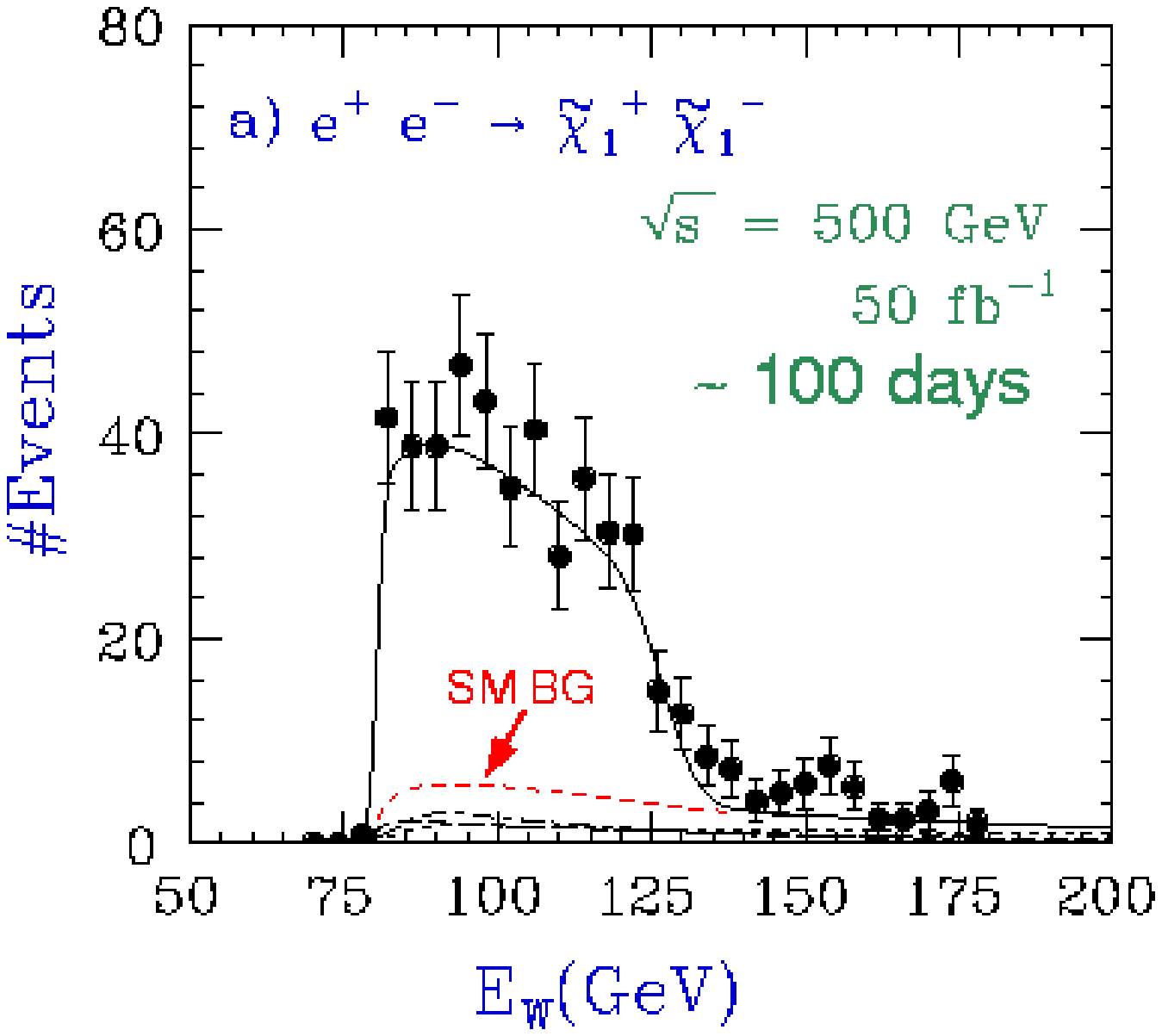}
}
\vspace {0.5in}
\centerline{
 \includegraphics*[scale=0.35]{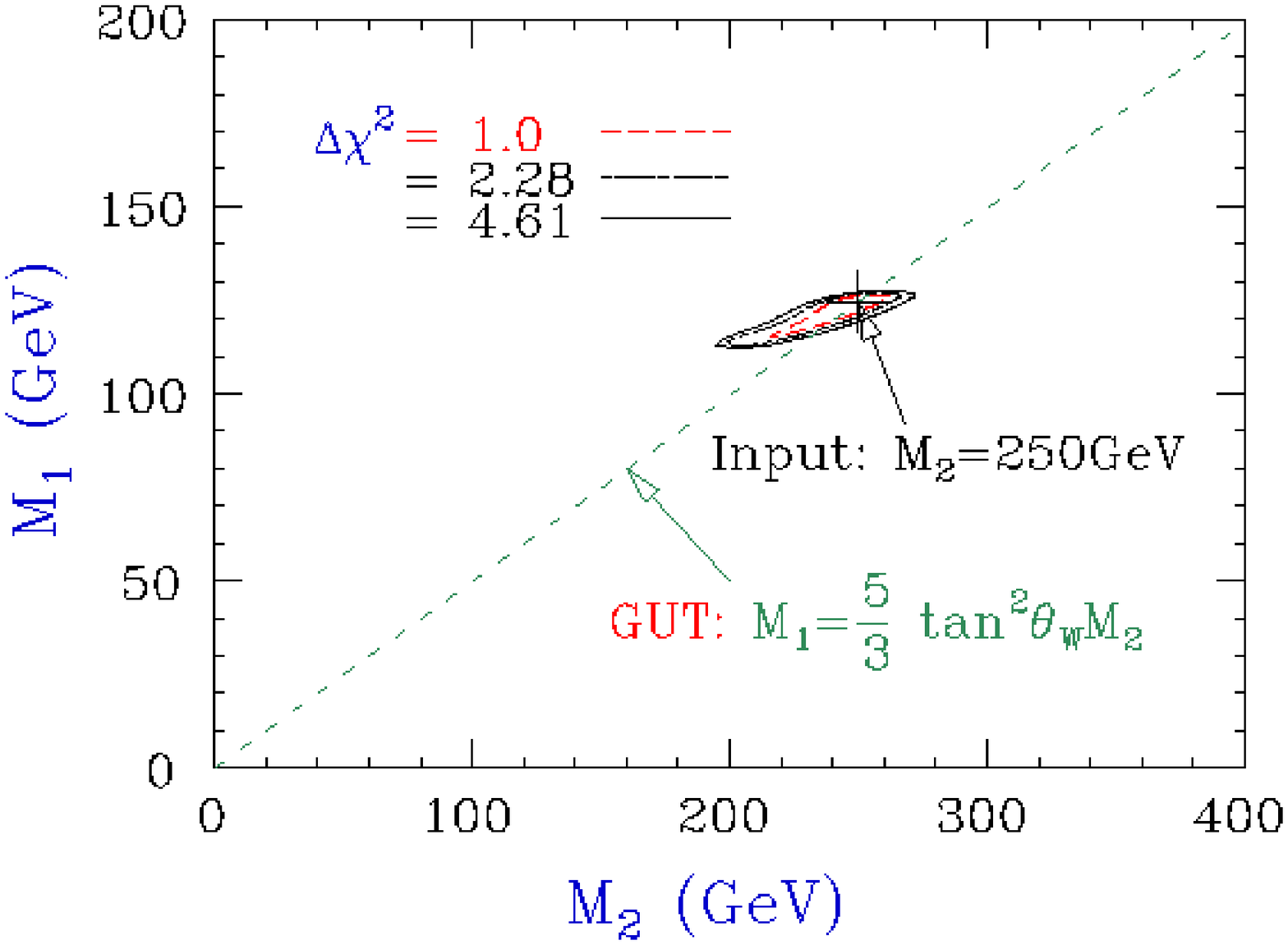}
}
\caption{Simultaneous determination of chargino/neutralino mass from 
chargino studies and consequent testing of the GUT relation between 
$M_1$ and $M_2$~\protect\cite{jlc2}.}
\label{F:rgplen:13}
\end{figure}
An example of this is
shown in Fig.~\ref{F:rgplen:13}. The right hand figure in the top panel 
shows how a determination of the energy distribution of the `W' produced in the
decay of $\Chip/\Chim $, in the reaction $\epem \to \Chip_1 \Chim_1$, affords a 
determination of $m_{\N0_1}$ and $m_{\Chipm}$ shown in the left panel. The
lower panel then shows how using the masses $m_{\Chipm},m_{\N0_1}$ along with
$\sigma_R(\Chimp_1 ), \sigma_R(\tilde e_R)$ and the angular distribution 
of the decay leptons one can extract $M_1,M_2$ at the GUT scale and test
the GUT relation.

A different approach~\cite{tesla} is to use the 
experimental observables such as 
cross-sections, angular distributions to determine the physical parameters
of the system such as masses and mixings and then use these to determine the
Lagrangian parameters $M_1,M_2,\mu,\tan \beta$ at the {\it EW scale itself}. 
Thus the possible errors of measurements of the experimental quantities alone 
will  control the accuracy of the detrmination of these parameters. There are
again two ways in which this information can be used: one is a top down 
approach which in spirit is similar to the earlier one as now one uses these
accurately determined Lagrangian parameters at the EW scale to fit their values
at the high scale and then compare them with the input value. 

\begin{figure}[htb]
\centerline{
 \includegraphics*[width=3.00in,height=2.00in]{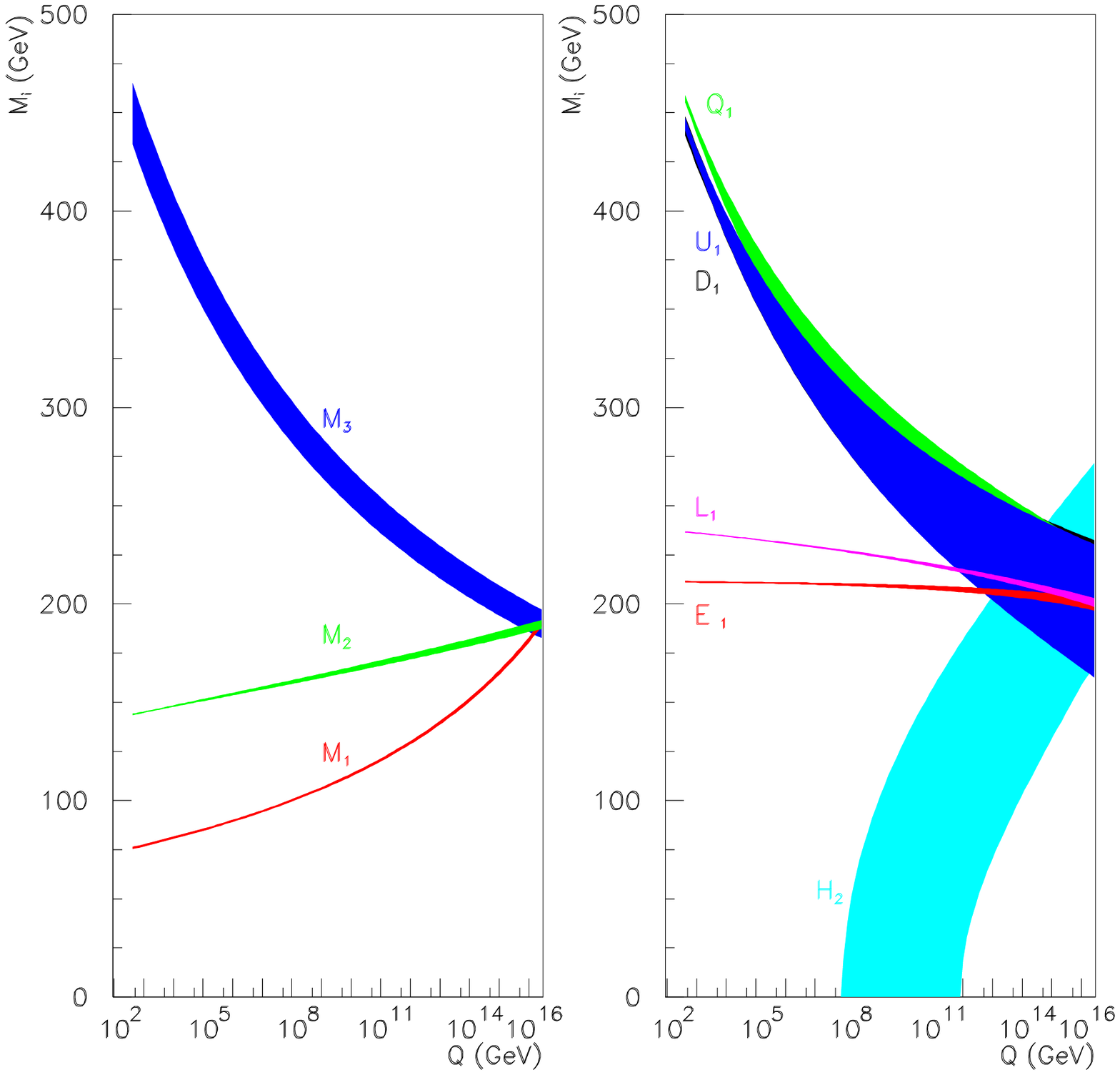}
 \includegraphics*[width=2.40in,height=2.00in]{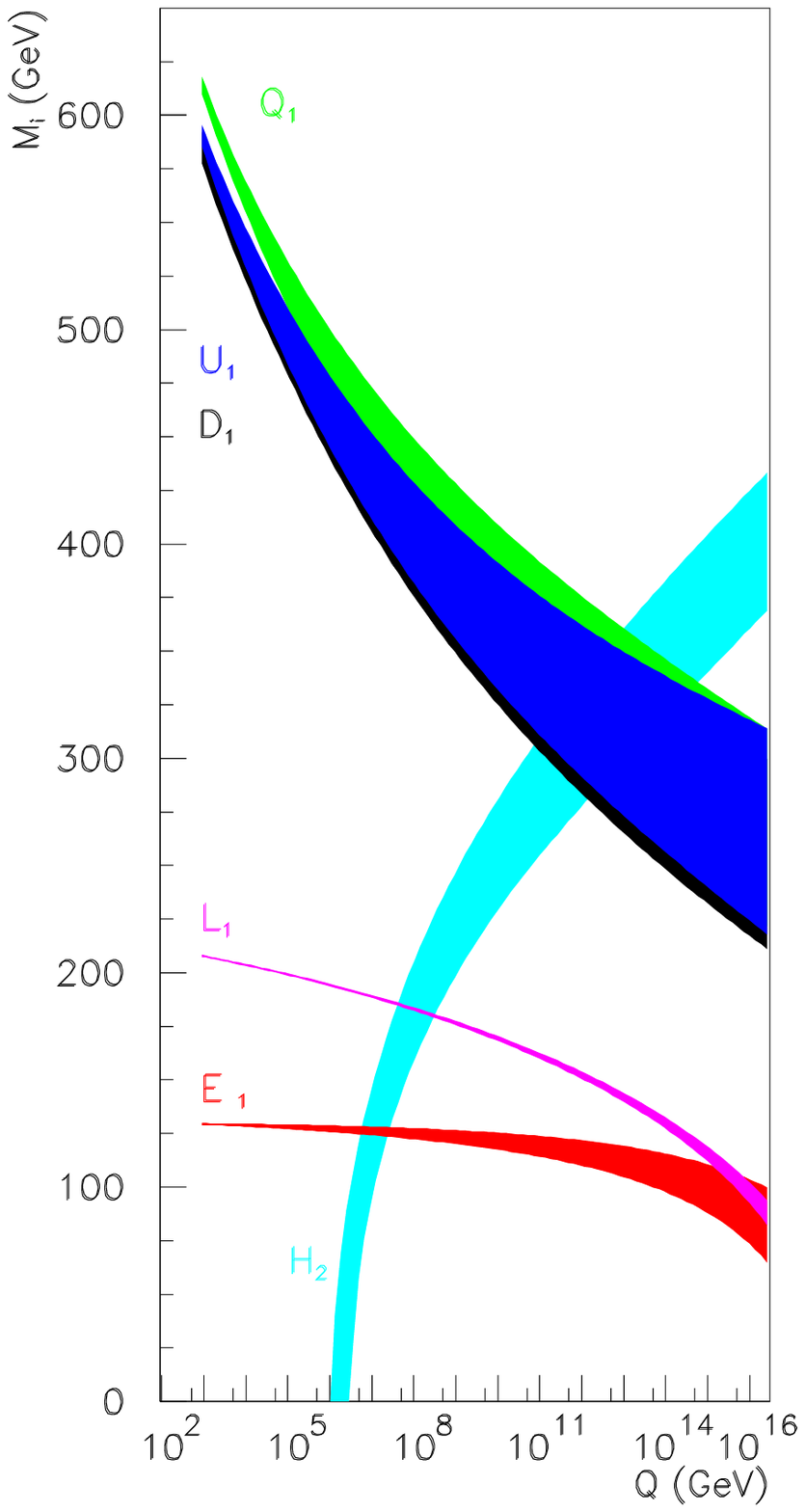}
}
\caption{Bottom up approach of the determination of the sparticle mass
parameters for mSUGRA and GMSB\protect\cite{PZB}. 
Values of the model parameters as given there.}
\label{F:rgplen:14}
\end{figure}

A completely different and a very interesting way of using the information on
these masses~\cite{PZB} is the bottom up approach where, one starts with these 
Lagrangian  parameters extracted at the weak scale and use the renormalisation 
group evolution (RGE) to calculate these parameters at the high scale. As 
explained in the introduction, different SUSY breaking mechanisms differ in 
their predictions for  relations among these various parameters at the high
scale. The ineteresting aspect of the bottom up approach is the possibility 
they offer of testing these relations `directly' by reconstructing them
from their low energy values using the RGE. In the analysis the
`experimental' values of the various sparticle masses are  generated in a
given scenario (mSUGRA,GMSB etc.) starting from the universal parameters
at the high scale appropriate for the model under consideration and using
the evolution from the high scale to the EW scale. These quantities are then
endowed with experimental errors expected to be reached in the combined 
analyses from LHC and an LC with energy upto $1$ TeV , with an integrated
luminosity of $1 {\rm ab}^{-1}$. Then these values are evloved once again
to the high scale. The  figure in the left panel of Fig.~\ref{F:rgplen:14}
shows  results of such an exercise for  the gaugino and sfermion masses for 
the mSUGRA case  and the one on the right for sfermion masses in GMSB. 
The width of the bands indicates $95 \%$ C.L.  Bear in mind 
that such accuracies will require a 10-20 year program at an LC 
with $\sqrt{s} \leq 1.5 $ TeV.
The two figures in the left panel show that with the projected 
accuracies of measurements, the unification of the gaugino masses will
be indeed demonstrated very clearly.

All these discussions assume that most of the sparticle spectrum
will be accessible jointly between the LHC and a TeV energy LC. 
If however, the squarks are superheavy~\cite{matchev,fengbagger}
(a possibility allowed by models) then perhaps the only
clue to their existence can be obtained through the analogue of precision
measurements of the oblique correction to the SM parameters at the Z pole. These
superoblique corrections~\cite{MIHO3}, modify the 
equalities between various couplings mentioned already in Eq.~\ref{eq:rg:1}.
These modifications arise if there is a large mass splitting between 
the sleptons and the squarks. The expected radiative corrections imply
$$
{{\delta g_Y} \over g_Y}  \simeq  {{11 g_Y^2} \over {48 \pi^2}} 
ln \left({m_{\sq}} \over {m_{\tilde l}}\right).
$$
Thus if the mass splitting is a factor 10 one expects a deviation from the
tree level relation by about $0.7 \% $. The discussions of the earlier
section demonstrate that it might be possible at an LC to make such a 
measurement. 

\section{Exploring the `SUSY'less option at Colliders}
In the introduction we saw that  apart from the
option of elementary higgs and SUSY there exist also two other
options to handle the hierarchy  problems; viz. the composite 
scalars~\cite{Morioka} and extra 'large' dimensions; warped or otherwise
~\cite{RS,ADD}. We argued that in the former case one will 
see evidence of  deviation of the trilnear and quartic  gauge boson 
couplings from the SM predictions.
\begin{figure*}[htbp]
\begin{center}
{\mbox{
\includegraphics[width=10cm,height=8cm]{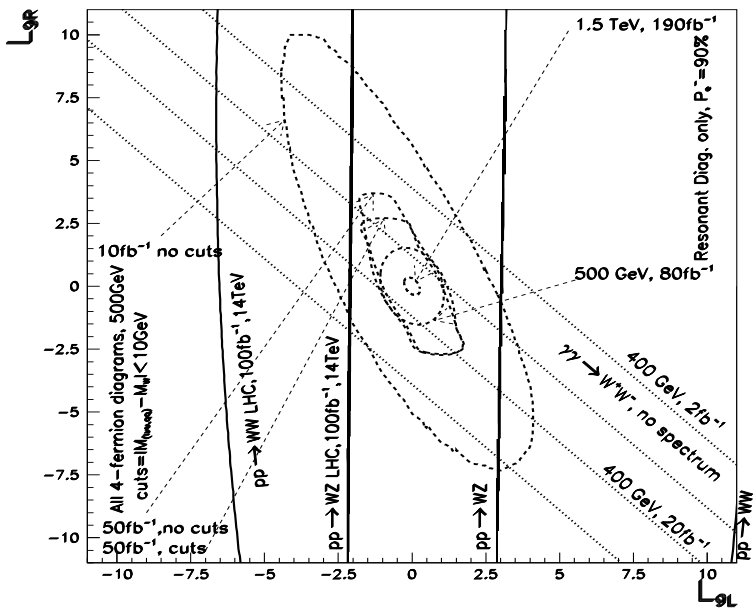}
}}
\end{center}
\caption{\label{l9limit}{\em Comparison of limits on the chiral
Lagrangian parameters $L_9$ at the future colliders\/,
from\cite{Morioka}.}}
\end{figure*}
Fig.~\ref{l9limit} taken from Ref.~\cite{Morioka} 
shows an example of the reach for the particular
operator $L_9$ at the LHC and the planned \eplem\, \gamgam\ colliders.

The whole development of the subject of `large' extra dimensions at LHC
is a very good example as to  how the various features of the 
detectors, 
such as good lepton detection, can be used very effectively in looking
for `new' physics which was {\bf not} taken into account while designing
the detector.  In the context of
LHC, the clearest signal for these `large' dimensions is via the observation
of graviton resonances\cite{hewett} in the dilepton spectrum via the process
$gg \to G \to l^+l^-$. It has been demonstrated by Hewett et al\cite{hewett} 
that by using the constraints already available from the dijet/dilepton
data from the Tevatron and making reasonable assumptions so that the EW scale
is free from hierarchy problem, in the scenario with `warped' extra 
dimesnions\cite{RS}, the parameter space of the model can be completely 
covered at LHC using the dilepton channel.

Apart from determining the mass of the graviton, it is also essntial to 
check the spin of the exchanged particle. ATLAS performed an 
analysis\cite{park}, which showed that the acceptance of the detector 
is quite low at large $\cos \theta^*$. 
\begin{figure}[htb]
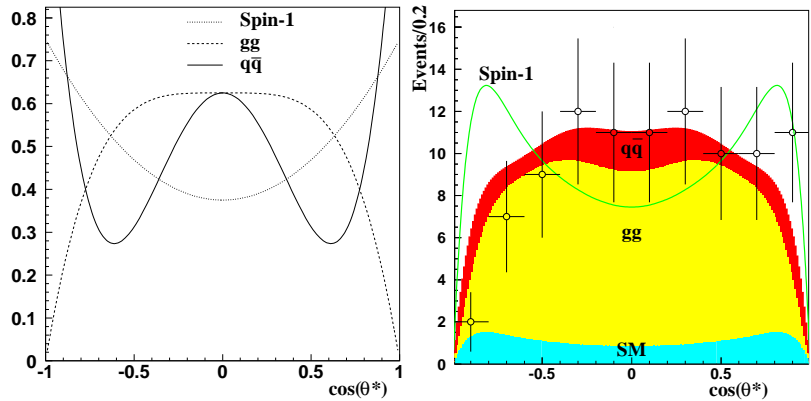

      \centerline{
      \includegraphics*[scale=0.3]{ang_theory.epsi}
      \includegraphics*[scale=0.3]{ang_experiment_1500.epsi}
  }
\caption{\em Angular distribution in the cm. frame for the $l^+l^-$ 
pair expected in the detector for the graviton along with the  expectation 
for a spin 1 particle at LHC\protect\cite{park}.}
\label{ften}
\end{figure}
The left panel of fig.~\ref{ften} shows the different angular distributions expected for
different spin exchanges. For the spin-2 case the contributions from the
$gg$ and $qq$ initial state are shown separately. The panel on the right 
shows that even with the lowered acceptance, it might be possible to 
discriminate against a spin-1 case.
Many more investigations on the subject are going on and the  end
conclusion is that it should be possible to see the effect of these
`large' compact dimensions (warped or otherwise) at the LHC upto 
almost all the values of the model parameters which seem reasonable and 
for which the theoretical formulation remains consistent.

The effects of these 'large' extra dimensions can be studied at the \eplem\
and \gamgam\ colliders very easily, the available polarisation serving a 
very useful role~\cite{appc1,sridhar,tesla,nlc,jlc}. Studies of large
$p_T$ jet production, $t \bar t$ pair production, fermion pair production 
can see these `extra'  large dimensions if they correpsond to the TeV scale
which should be the scale at which they should appear if they have to obviate
the need of SUSY to solve the hierarchy problem. The joint reach of the
LHC and the next linear colliders will certainly cover that range.

\section{Conclusion}

In conclusion we can say that the TeV energy colliders, both hadronic and 
leptonic, are necessary to further  our understanding of the fundamental 
particles and interactions among them and that such colliders will definitely
be able to provide further answers to this very basic question that we ask.
Such colliders should 1)either be able to find a fundamental higgs scalar,  
study its properties in detail to test whether it is the SM higgs  
and 2) obtain evidence for  existence of Supersymmetry (which seems 
necessary for the theoretical consistency of the SM) if nature has 
chosen the supersymmetric path  and obtain 
information about the Supersymmetry breaking and breaking scale. 
3) Alternatively they should be able to find evidence for some
other physics beyond the SM such as composite scalars or the
extra `large' dimensions which do not need the existence of supersymmetry
to solve the hierarchy problem.
In any situation the physics prospects of the currently running colliders 
like the Tevatron, future  collider like the LHC which is now under preparation
and the linear \eplem , \gamgam\ colliders which are in the planning stages
are extremely exciting and hold a lot of promise to help us understand some 
very basic issues about the elementary  particles and their interactions.

\end{document}